\documentclass[12pt]{extarticle}
\usepackage[english]{babel}
\usepackage[margin=1in, letterpaper]{geometry}
\usepackage[english]{babel}
\usepackage[utf8x]{inputenc}
\usepackage[T1]{fontenc} 
\usepackage{amsmath}
\usepackage{graphicx}
\usepackage[colorinlistoftodos,prependcaption,textsize=small]{todonotes}
\usepackage{xargs}                      
\usepackage{filecontents}
\usepackage{natbib, bibentry}		
\usepackage[pdfencoding=unicode]{hyperref}
\usepackage{parskip} 
\usepackage{colortbl}
\usepackage{tabularx}
\usepackage{titlesec}
\usepackage[titletoc, title]{appendix}
\usepackage{subfigure}
\usepackage{boldline, multirow}
\usepackage{array}
\usepackage{arydshln}
\usepackage{cprotect}
\usepackage{adjustbox}
\usepackage{pdflscape}
\usepackage{times}
\usepackage{titlesec}
\usepackage{fancyhdr}
\usepackage[hang,flushmargin,perpage]{footmisc}
\usepackage[normalem]{ulem}
\usepackage{nicefrac}
\usepackage{amsbsy}
\usepackage{afterpage}
\usepackage{enumitem}		
\usepackage{datetime}		
\usepackage{amssymb} 		
\usepackage{makecell} 		
\usepackage{hhline} 			
\usepackage{lineno}			
\usepackage{relsize}			
\usepackage[skip=0.5\baselineskip]{caption}		
\usepackage{amsbsy}

\usepackage[titletoc, title]{appendix}

\newdateformat{monthyeardate}{
  \monthname[\THEMONTH] \THEYEAR}

\usepackage[nodisplayskipstretch]{setspace}		

\hypersetup{
    colorlinks=true,
    linkcolor=blue,
    citecolor=blue,
    filecolor=magenta,
    urlcolor=blue,
}

\bibpunct{(}{)}{,}{a}{}{,~}

\titleformat{\section}
  {\normalfont\fontsize{14}{14}\bfseries}{\thesection}{1em}{}

\titleformat{\paragraph}
{\normalfont\fontsize{12}{12}\bfseries}{\theparagraph}{1em}{}
\titlespacing*{\paragraph}
{0pt}{3.25ex plus 1ex minus .2ex}{1.5ex plus .2ex}

\newcommandx{\unsure}[2][1=]{\todo[linecolor=red,backgroundcolor=red!25,bordercolor=red,#1]{#2}}
\newcommandx{\change}[2][1=]{\todo[linecolor=blue,backgroundcolor=blue!25,bordercolor=blue,#1]{#2}}
\newcommandx{\info}[2][1=]{\todo[linecolor=green,backgroundcolor=green!25,bordercolor=green,#1]{#2}}
\newcommandx{\improvement}[2][1=]{\todo[linecolor=pink,backgroundcolor=pink!25,bordercolor=pink,#1]{#2}}
\newcommandx{\thiswillnotshow}[2][1=]{\todo[disable,#1]{#2}}
\newcommandx{\checkthat}[2][1=]{\todo[linecolor=gray,backgroundcolor=gray!25,bordercolor=gray,#1]{#2}}

\usepackage{soul,color}

\presetkeys{todonotes}{size=\setstretch{0.8}\selectfont}{}

\newcommand{\0}{\text{0}}
\newcommand{\1}{\text{1}}
\newcommand{\2}{\text{2}}
\newcommand{\3}{\text{3}}
\newcommand{\4}{\text{4}}
\newcommand{\comma}{\text{,}}

\reversemarginpar


\newcolumntype{L}[1]{>{\raggedright\arraybackslash}p{#1}}  
\newcolumntype{C}[1]{>{\centering\arraybackslash}p{#1}} 
\newcolumntype{R}[1]{>{\raggedleft\arraybackslash}p{#1}} 
\newcolumntype{Y}{>{\centering\arraybackslash}X}

\usepackage{accents}

\usepackage{rotating}					
\title{ \Large \textbf{Pulse-Period -- Moment-Magnitude Relations\\ \vspace*{-0.3cm}Derived with Wavelet Analysis and their Relevance\\ \vspace*{-0.3cm}to Estimate Structural Deformations}}
\author{\vspace{-5ex}}
\date{\vspace{-5ex}}

\makeatletter
\renewenvironment{abstract}{%
    \if@twocolumn
      \section*{\abstractname}%
    \else 
      \begin{center}%
        {\bfseries \large\abstractname\vspace{-0.7cm}}
      \end{center}%
      \quotation
    \fi}
    {\if@twocolumn\else\endquotation\fi}
\makeatother

\begin{document}


\maketitle

\vspace*{-1.8cm}

\begin{center}
{\large Eleftheria Efthymiou$^\text{1}$ and Nicos Makris$^\text{1,2,*}$}

$^\text{1}$\textit{Dept. of Civil \& Environmental Engineering, Southern Methodist University, Dallas, TX, 75276} \vspace*{0.2cm} \\
$^\text{2}$\textit{Office of Theoretical and Applied Mechanics, Academy of Athens, 106 79, Greece} \\ \vspace*{0.15cm}
$^\text{*}$nmakris@smu.edu
\end{center}

\vspace*{0.05cm}
\begin{abstract}
\singlespacing
{\small %
Motivated from the quadratic dependence of peak structural displacements to the pulse period, $T_p$, of pulse-like ground motions, this paper revisits the pulse-period -- moment-magnitude ($T_p$--$M_\text{W}$) relations of ground motions generated from near-source  earthquakes with epicentral distances, $D\leq$ 20 km. A total of 1260 ground motions are interrogated with wavelet analysis to identify energetic acceleration pulses (not velocity pulses) and extract their optimal period, $T_p$, amplitude, $a_p$, phase, $\phi$ and number of half-cycles, $\gamma$. The interrogation of acceleration records with wavelet analysis is capable of extracting shorter-duration distinguishable pulses with engineering significance, which override the longer near-source pulses and they are not necessarily of random character. Our wavelet analysis identified 109 pulse-like records from normal faults, 188 pulse-like records from reverse faults and 125 pulse-like records from strike-slip faults, all with epicentral distances $D\leq$ 20 km. Regression analysis on the extracted data concluded that the same $T_p$--$M_\text{W}$ relation can be used for pulse-like ground motions generated either from strike-slip faults or from dip-slip normal faults; whereas, a different $T_p$--$M_{\text{W}}$ relation is proposed for dip-slip reverse faults. The study concludes that for the same moment magnitude, $M_{\text{W}}$, the pulse periods of ground motions generated from strike-slip faults are on average larger than these from reverse faults --- a result that is in agreement with findings from past investigations. Most importantly, our wavelet analysis on acceleration records produces $T_p$--$M_{\text{W}}$ relations with a slope that is lower than the slopes of the $T_p$--$M_{\text{W}}$ relations presented by past investigators after merely fitting velocity pulses. As a result, our proposed $T_p$--$M_{\text{W}}$ relations yield lower $T_p$ values for larger-magnitude earthquakes (say $M_{\text{W}}>$ 6), allowing for the estimation of dependable peak structural displacements that scale invariably with $a_pT_p^{\text{2}}$.}
\onehalfspacing
\end{abstract}


\section{Introduction}\label{sec:Sec01}
\vspace*{-0.5cm}
One of the important  challenges in earthquake engineering is the estimation of peak inelastic structural displacements for collapse prevention and avoidance of irreparable damage. Traditionally, earthquake shaking was viewed as a random motion characterized essentially by two parameters: (a) its peak ground acceleration (PGA); and (b) its overall duration, $T_D$, and this led to the worldwide accepted equivalent static lateral force procedure \citep{UBC1997,FEMA273_1997,IBC2000,FEMA2000}. Fourier spectra of excitations have been shown occasionally in technical publications; yet, the frequency content of ground motions was a concern only in the design of special projects. The random character of earthquake excitations was bypassed in engineering design by establishing the design elastic response spectrum with a horizontal constant value from low to medium periods and a descending tail at longer periods, so that any candidate seismic motion used for the structural response analysis at a given site has to be compatible to the extent possible with the proposed design spectrum.

\begin{figure}[t!]
\centering
  \includegraphics[width=.75\linewidth]{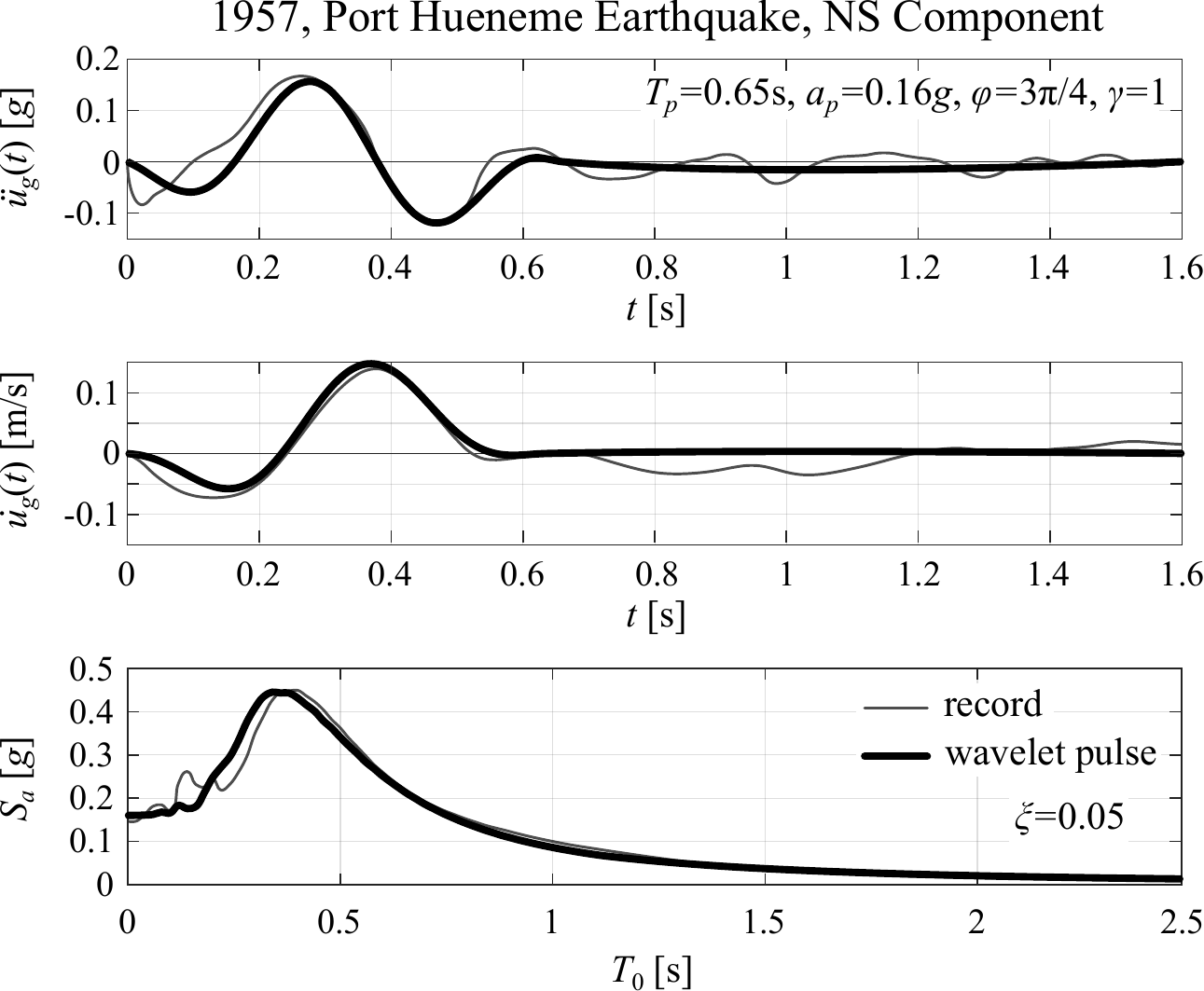}
\caption{Acceleration and velocity time histories together with the associated elastic acceleration response spectra of the North--South record from the March 18, 1957 Port Hueneme, California earthquake (thin solid lines) and of a wavelet pulse that best fits the acceleration record (heavy solid lines \citep{VassiliouMakris2011}).}
\label{fig:Fig01}
\end{figure}

The strong motion recorded by the United States Coast and Geodetic Survey (USCGS) accelerograph at Port Hueneme, California on March 18, 1957 was the first recorded strong seismic motion which consisted essentially of a single pulse \citep{HousnerHudson1958}. Figure \ref{fig:Fig01} plots the North--South component of the ground motion recorded during the 1957 Port Hueneme earthquake (thin solid lines) together with the best fit of a mathematical wavelet \citep{VassiliouMakris2011} on the acceleration record with a pulse period $T_p=$ 0.65 s and an acceleration pulse amplitude, $a_p=$ 0.16$g$ (heavy solid lines). In their insightful paper \cite{HousnerHudson1958} indicate that the elastic response spectrum values from the 1957 Port Hueneme record (also shown in Fig. \ref{fig:Fig01} (bottom)) are considerably larger than the values from the spectra of most typical Pacific Coast earthquakes of equivalent magnitude. These abnormally high-values were reflected in an unusual amount of structural damage from an earthquake of magnitude 4.7 \citep{HousnerHudson1958}. 

The USCGS strong motion accelerometer was approximately 5 miles northwest of the earthquake epicenter, and the 1957 Port Hueneme record marks the beginning of a series of investigations on the pulse-like nature of near-source ground motions \citep{Bolt1971, Bolt1975} and the severe inelastic displacement demand on structures that result from the coherent, long-duration acceleration pulses \citep{BerteroHerreraMahin1976,BerteroMahinHerrera1978,BerteroAndersonKrawinklerMiranda1991,SomervilleGraves1993}. Such investigations were drastically intensified after the 1994 Northridge, California, the 1995 Kobe, Japan, the 1999 Izmit, Turkey and the 1999 Chi-Chi, Taiwan earthquakes which resulted in a large collection of pulse-like ground motions recorded near the causative faults \citep[and references reported therein]{HallHeatonHallingWald1995,Makris1997,LohLeeWuPeng2000,MakrisChang2000,MaMoriLeeYu2001, WangChangChen2001,SekiguchiIwata2002}.

\begin{figure}[t!]
\centering
  \includegraphics[width=.75\linewidth]{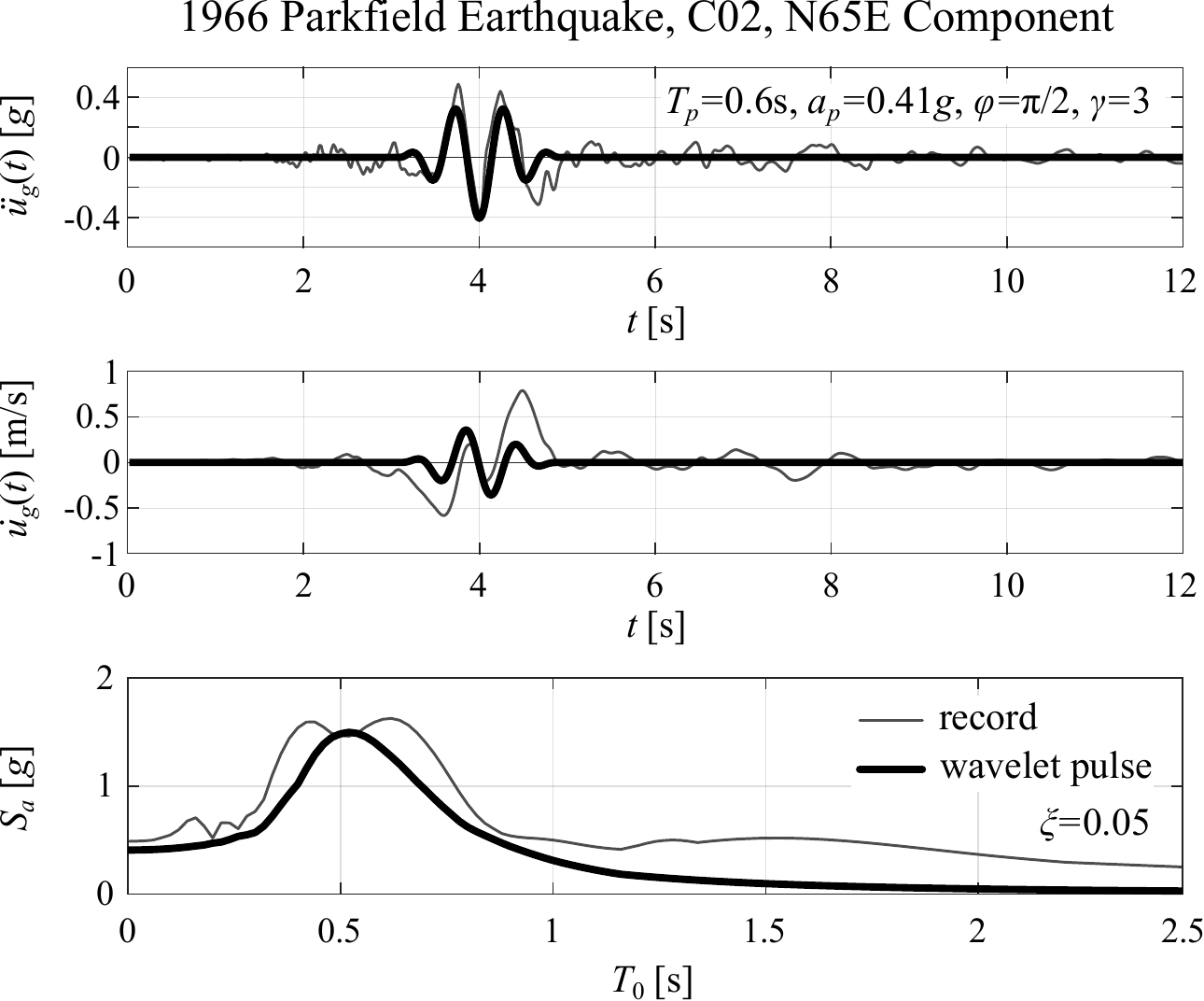}
\caption{Acceleration and velocity time histories together with the associated elastic acceleration response spectra of the N65E component of the C02 record from the June 28, 1966 Parkfield, California earthquake(thin solid lines) and of a wavelet pulse that best fits the acceleration record (heavy solid lines \citep{VassiliouMakris2011}).}
\label{fig:Fig02}
\end{figure}

Less than a decade after the \cite{HousnerHudson1958} seminal paper, the $M_{\text{W}}=$ 6.2, 1966 Parkfield, California earthquake generated ground shaking that was recorded by an array of five strong-motion accelerographs across the San Andreas fault at Cholame. Among the records, the most notable one is the C02 record at a distance of only 80 m from the fault rupture. Figure \ref{fig:Fig02} plots the N65E component of the C02 record together with the best fit of a mathematical wavelet \citep{VassiliouMakris2011} on the acceleration record. The response spectra for this motion were first presented by \cite{HousnerTrifunac1967} (also shown at the bottom of Fig. \ref{fig:Fig02}) who indicated that in an engineering sense the 1966 Parkfield ground motion belongs to a different class than the other strong ground motions recorded before that time; given the $T_p=$ 0.6-s long coherent acceleration pulse that dominates the acceleration record.

\begin{figure}[t!]
\centering
  \includegraphics[width=\linewidth]{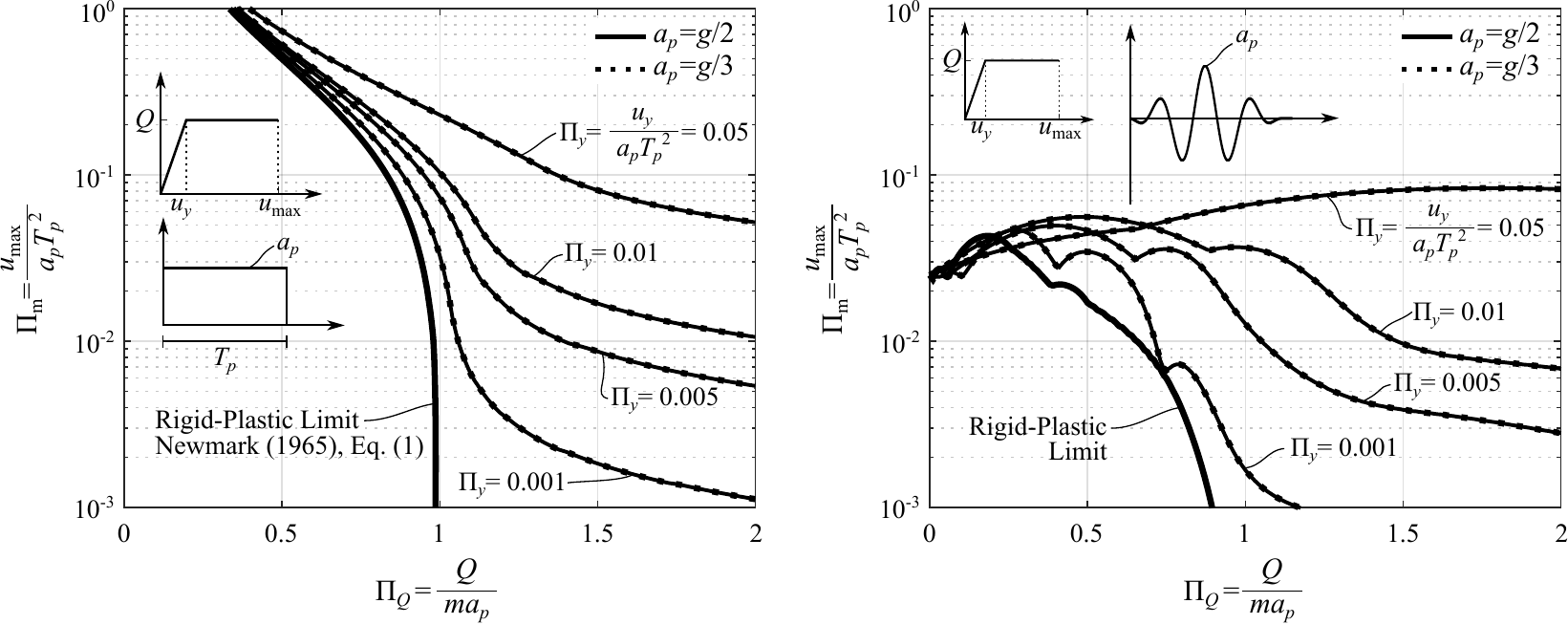}
\caption{Normalized maximum relative displacement curves $\dfrac{u_{\text{max}}}{a_p T_p^\2}$ of an elastic--plastic structure subjected to a rectangular pulse (left) and a M\&P wavelet pulse with $\phi=\pi/\text{2}$ and $\gamma=$ 3 (right). The self-similar solutions approach the rigid--plastic limit as the normalized yield displacement $\Pi_y=\dfrac{u_y}{a_p T_p^\2}$ tends to zero.}
\label{fig:Fig03}
\end{figure}

The first systematic work on the response of an elastoplastic single-degree-of-freedom (SDoF) structure subjected to pulse-type ground shaking was presented in the papers by \cite{VeletsosNewmark1960} and \cite{VeletsosNewmarkChelapati1965}. An early solution to the response of a rigid--plastic system (rigid mass sliding on its moving base with coefficient of friction, $\mu$) subjected to a rectangular acceleration pulse with amplitude $a_p$ and duration $T_p$ was presented in \citeauthor{Newmark1965}'s (1965) Rankine Lecture who showed that the maximum sliding displacement of a mass relative to its shaking base is
\begin{equation}\label{eq:Eq01}
u_{\text{max}}=\frac{a_p T_p^\2}{\2}\left( \frac{a_p}{\mu g}-\1  \right)\comma \quad a_p > \mu g
\end{equation}
Equation \eqref{eq:Eq01} indicates that the maximum sliding (inelastic) displacement is proportional to the pulse acceleration $a_p$ (that is the PGA for a rectangulat pulse); and most importantly, is proportional to the square of the duration of the pulse (pulse period) $T_p^\2$. The product $a_p T_p^\2=L_p$ in front of the parenthesis of Eq. \eqref{eq:Eq01} is a characteristic length scale that expresses the persistence of the pulse to induce inelastic deformations; whereas, the quantity in the parenthesis $\dfrac{a_p}{\mu g}$ expresses the strength of the pulse. While the rectangular pulse used by \cite{Newmark1965} is not physically realizable by earthquake shaking (results to an infinite ground displacement) it shows in a lucid manner that inelastic deformations are proportional to the square of the duration of the pulse $T_p^\2$ --- a physical reality that has been overlooked invariably by seismic codes worldwide for several decades. The scaling of the maximum sliding displacement with $T_p^\2$ shown by Eq. \eqref{eq:Eq01} was obtained by \cite{Newmark1965} after integrating the differential equation of motion of the simple rigid-plastic system. The same result can be obtained with dimensional analysis for the response of any given structural system subjected to any pulse-like excitation \citep{MakrisBlack2004Dimensional_a,MakrisBlack2004Dimensional_b} such as the pulse motions shown in Figs. \ref{fig:Fig01} and \ref{fig:Fig02}. Figure \ref{fig:Fig03} plots the normalized maximum inelastic displacement $\Pi_\text{m}=\dfrac{u_{\text{max}}}{a_p T_p^\2}$ of an elastoplastic structure subjected to a rectangular pulse (left) and the symmetric acceleration wavelet ($\phi=\pi/\text{2}$, $\gamma=$ 3) that happens to best match the C02/NE65 acceleration record shown in Fig. \ref{fig:Fig02} (right) as a function of the dmensionless strength $\Pi_Q=\dfrac{Q}{ma_p}$ for various values of the dimensionless yield displacement $\Pi_y=\dfrac{u_y}{a_p T_p^\2}$. The self-similar solutions approach the rigid-plastic limit as the normalized yield displacement $\Pi_y=\dfrac{u_y}{a_p T_p^\2}$ tends to zero \citep{MakrisBlack2004Dimensional_a}. Figure \ref{fig:Fig03} uncovers that the non-reversing rectangular pulse induces much larger displacements than the wavelet that happens to best match the C02 record shown in Fig. \ref{fig:Fig02} as the strength of the structure reduces.

The scaling of the peak displacement $u_{\text{max}}$, with $T_p^\2$ derived by \cite{Newmark1965} for a sliding mass is independent of the structural behavior as illustrated in Fig. \ref{fig:Fig03}. For instance, in the interest of completeness, consider an elastic SDoF oscillator with mass $m$, stiffness $k=m\omega_\0^\2=m\dfrac{\4\pi^\2}{T_\0^\2}$ and damping constant $c=\2\xi m\omega_\0$ that is subjected to a given acceleration pulse with duration $T_p$ and acceleration amplitude $a_p$. The maximum relative-to-the-ground displacement of the elastic SDoF oscillator, $u_{\text{max}}$, is a function of four variables:
\begin{equation}\label{eq:Eq02}
u_{\text{max}}=f\left( T_\0 \comma\, \xi \comma\, a_p \comma\, T_p \right)
\end{equation}
where $T_\0=\dfrac{\2\pi}{\omega_\0}=\2\pi\sqrt{\dfrac{m}{k}}$ and $\xi=\dfrac{c}{\2 m\omega_\0}$ are the parameters of the elastic structure and $a_p$ and $T_p$ are the parameters of the pulse excitation. The five variables appearing in Eq. \eqref{eq:Eq02}, $u_{\text{max}}\doteq [\text{L}]$, $T_\0\doteq [\text{T}]$, $\xi\doteq [\1]$, $a_p\doteq [\text{L}][\text{T}]^{-\2}$ and $T_p\doteq [\text{T}]$, involve only two reference dimensions, that is length $[\text{L}]$ and time $[\text{T}]$. According to the Vaschy -- Buckingham $\Pi$ theorem \citep{Langhaar1951,HousnerHudson1959,Barenblatt1996}, the number of independent dimensionless products ($\Pi$-terms) is equal to the number of physical variables appearing in Eq. \eqref{eq:Eq02} $($5 physical variables$)$ minus the number of reference dimensions $($2 reference dimensions $[\text{L}] \text{ and } [\text{T}])$ of the physical problem. Accordingly, for the linear SDoF we have 5$-\2=$ 3 dimensionless $\Pi$-terms and Eq. \eqref{eq:Eq02} condenses to:
\begin{equation}\label{eq:Eq03}
\Pi_m=\frac{u_{\text{max}}}{a_p T_p^\2}=\phi\left( \frac{T_\0}{T_p} \comma\, \xi \right) \quad \text{or} \quad u_{\text{max}}= a_p T_p^\2 \phi\left( \frac{T_\0}{T_p} \comma\, \xi \right) 
\end{equation}
Equations \eqref{eq:Eq01} and \eqref{eq:Eq03} together with Fig. \ref{fig:Fig03} conclude to the same central result: that regardless whether the structural behavior is rigid-plastic (sliding mass), elastoplastic or elastic, the peak response displacement of any structural system when subjected to a coherent acceleration pulse with duration $T_p$ is proportional to $T_p^\2$. This result derives directly from dimensional analysis \citep{MakrisBlack2004Dimensional_a} and has been confirmed numerically for reinforced concrete structures \citep{MakrisPsychogios2006} and steel structures \citep{KaravasilisMakrisBazeosBeskos2010}. Consequently, given that the elastic or inelastic peak structural displacement scales invariably with $T_p^\2$, the aim of this paper is to produce with wavelet analysis, mathematically-objective $T_p$--$M_\text{W}$ relations after examining their possible dependence on the fault mechanism in an effort to obtain dependable estimates of pulse periods, $T_p$ so that Eq. \eqref{eq:Eq03} or the results of Fig. \ref{fig:Fig03} can be used with confidence to estimate structural deformations.

\section{Review of Existing $\pmb{T_p}$ -- $\pmb{M_\text{W}}$ Relations and Problem Statement}\label{sec:Sec02}
\vspace{-0.5cm}

A number of models relating the period of the coherent velocity pulse, $T_p$, to the moment magnitude of the earthquake, $M_\text{W}$ are available in the literature. \citet{Somerville1998} examined the period and amplitude of the largest cycle of motion of the fault-normal, forward directivity velocity pulse. This pulse consists of a peak, a trough and three zero crossings. With this idealization, the pulse amplitude is equal to the peak ground velocity (\textit{PGV}) and the pulse period is equal to the duration of the velocity cycle. The proposed model relates the pulse period, $T_p$, with the earthquake's moment magnitude, $M_\text{W}$, assuming that the period is independent of the distance:
\begin{equation}\label{eq:Eq04}
\log_{\text{10}}T_p = -\text{2.5}+\text{0.425}M_\text{W}
\end{equation}
By constraining the model to be self-similar, \citet{Somerville1998} obtained:
\begin{equation}\label{eq:Eq05}
\log_{\text{10}}T_p = -\text{3.0}+\text{0.50}M_\text{W}.
\end{equation}
\cite{Somerville1998} used records that presented forward directivity only, which are known to produce strong pulses in the fault-normal direction. 15 time histories recorded in close proximity to the earthquakes epicenters: 0--10 km with magnitudes, $M_\text{W}$, between 6.2 and 7.3 were used. The database was complimented by 12 simulated time histories in distance of 3 to 10 km and magnitudes of 6.5 to 7.5. All time-histories are for soil site conditions. A linear relationship between the pulse period $T_p$ and the rise time of slip on the fault, $T_R$, was identified which is a measure of the duration of slip at a specific point of the fault: $T_p=\text{2.2}T_R$. For a given $M_\text{W}$ the rise time in a reverse-faulting earthquake is on average about half that of strike-slip earthquakes, suggesting that the larger ground motion levels of reverse-faulting earthquakes can be attributed to the shorter rise times than strike-slip earthquakes; therefore, the pulse periods, $T_p$, of pulse-like ground motions from reverse faults are expected to be of shorter duration than the pulse periods of ground motions from strike-slip faults.

\citet{KrawinklerAlavi1998} explored the elastic and inelastic dynamic responses of SDoF and MDoF systems subjected to different types of pulse loading. They proposed three different acceleration pulses defined in terms of their period, $T_p$, and their peak ground acceleration, $a_{g\text{, }\textit{max}}$. The non-differentiable square pulses denoted P1, P2 and P3 correspond to a half-pulse, a full pulse and a multiple pulse, respectively. Pulse P1 is half the P2 pulse. \citet{AlaviKrawinkler2000} later used the proposed full pulse model P2 to obtain the pulse characteristics of a small set of recorded near-fault ground motions. With regression analysis of the pulse periods of the velocity pulses against the earthquake moment magnitude they obtained:
\begin{equation}\label{eq:Eq06}
\log_{\text{10}}T_p = -\text{1.76}+\text{0.31}M_\text{W}
\end{equation}
where $T_p$ is the pulse period, defined as the duration of the complete velocity cycle and $M_\text{W}$ is the moment magnitude of the earthquake. \citet{AlaviKrawinkler2000} note that the records used in the regression analysis came from different faulting mechanisms and various geology conditions. They also warned that the lack of adequate numbers of recorded near-fault motions suggests that such relationships must be used with caution. However, by cautiously using the proposed relationships and combined with pulse strength demand spectra, one can evaluate the base shear strength required to limit story ductility ratios to specific target values.

In a subsequent paper, \citet{Somerville2003} used a triangular pulse that resembles the P2 velocity pulse of \citet{KrawinklerAlavi1998} and proposed two additional relationships that relate the period of the largest cycle of the fault-normal velocity waveform that presents forward directivity, $T_p$, with $M_\text{W}$. The expressions proposed for rock and soil sites assume that the period is independent of the distance from the fault:
\begin{align}\label{eq:Eq07}
\log_{\text{10}}T_p & =
\begin{cases}
 -\text{3.17}+\text{0.5}M_\text{W} \text{,} \quad \text{for rock sites} \\
 -\text{2.02}+\text{0.346}M_\text{W} \text{,} \quad \text{for soil sites}
\end{cases}
\end{align}
For the rock sites \citet{Somerville2003} proposed a self-similar expression, but for soil sites the expression is allowed to depart from self-similarity to accommodate effects from nonlinear soil response. The soil layers tend to increase the peak velocity and the period of the input rock motion by factors that depend on the magnitude of the ground motion and on the thickness and the properties of the soil layers \citep{RodriguezMarek2000, Somerville2003}. The two expressions intersect at $M_\text{W}=$ 7.4 and $T_p=$ 3.4 s. Above this point, the relationship for rock sites results in larger $T_p$ than the one for soil sites, although the author states that it is expected that the expression for soil sites would be curved and merge with the expression for rock sites above the intersection point. 

\citet{BrayRodriguezMarek2004} examined approximately 50 records presenting forward directivity and proposed a linear relation between $\ln T_p$ and the moment magnitude, $M_\text{W}$ after adopting sine pulses to represent simplified velocity-time histories. They distinguished between the fault normal and fault parallel directions of propagation by introducing the time-lag between the initiation of the fault normal and the fault parallel components. They accepted the number of significant pulses as the number of half-cycle velocity pulses that have amplitudes of at least 50\% of the \textit{PGV} of the ground motion. By complementing  the work of \citet{Somerville1998}, the \citeauthor{BrayRodriguezMarek2004} work accounts for distances smaller than 3 km from the fault, which were shown to be of significance, and proposes equations for rock sites, soil sites and for the complete database:
\begin{align}\label{eq:Eq08}
\ln T_p & = 
\begin{cases}
-\text{6.37}+\text{1.03}M_\text{W} \text{,} \quad \text{for all sites} \\ 
-\text{8.60}+\text{1.32}M_\text{W} \text{,} \quad \text{for rock sites} \\ 
-\text{5.60}+\text{0.93}M_\text{W} \text{,} \quad \text{for soil sites}  
\end{cases}
\end{align}
where $T_p$ is the period of the dominant velocity pulse. The regression is constrained to render same periods for rock and soil sites at $M_\text{W}=$ 7.6 ($T_p=$ 4.34 s)  in order to avoid obtaining larger periods for rock than for soil when $M_\text{W}>$ 7. 

\citet{MavroeidisPapageorgiou2003} indicated that simplified velocity pulses of square or triangular shapes do not adequately capture the time-histories or the response spectra of actual ground motions and their use to study the dynamic responses of structures may result in misleading conclusions. Their work builds on Gabor's (\citeyear{Gabor1946}) elementary signal, which is essentially a harmonic oscillation within a Gaussian envelope:
\begin{equation}\label{eq:Eq09}
v(t) = \exp \left[-\left(\dfrac{\text{2}\pi f_p}{\gamma} \right)^{\text{2}}t^{\text{2}}\right]\cos(\text{2}\pi f_p t + \nu)
\end{equation}
where $f_p$ is the prevailing frequency of the wavelet, $\nu$ is the phase angle and $\gamma$ controls its oscillatory character. \citet{MavroeidisPapageorgiou2003} showed that the Gabor wavelet of Eq. \eqref{eq:Eq09} does not allow closed-form solutions for the response of SDoF systems due to the exponential term in the right-hand side of Eq. \eqref{eq:Eq09}. In the interest of deriving closed-form expressions, they modified Eq. \eqref{eq:Eq09} by combining the harmonic oscillation part of the Gabor wavelet \citep{Gabor1946}, $\cos(\text{2}\pi f_p t + \nu)$, with an elevated cosine function of the form:
\begin{align}\label{eq:Eq10}
 v(t) & =
  \begin{cases} 
A \enskip \dfrac{\text{1}}{\text{2}}\Bigg[ \text{1} + \cos\Bigg( \dfrac{\text{2}\pi f_p}{\gamma} (t-t_{\text{0}}) \Bigg) \Bigg] \cos [ \text{2}\pi f_p (t-t_{\text{0}})+\nu]\text{,} \quad \quad \quad \quad \quad \quad \quad \quad\\
\text{0,} \quad \text{otherwise}
  \end{cases} \\
  & \quad \quad \quad \quad \quad \quad \quad \quad \quad \quad \quad \quad t_{\text{0}} - \dfrac{\gamma}{\text{2}f_p} \leq t \leq  t_{\text{0}} + \dfrac{\gamma}{\text{2}f_p} \quad \text{with } \gamma>\text{1} \text{ and } f_p = \dfrac{\text{1}}{T_p} \nonumber
\end{align}
where $A$ and $f_p=\text{1}/T_p$ denote the amplitude and the frequency of the velocity pulse, respectively, $\nu$ is the phase of the pulse, $\gamma$ defines the oscillatory character of the pulse and $t_{\text{0}}$ defines the time instance when the wavelet is located. The proposed velocity expression of Eq. \eqref{eq:Eq10} can capture effectively the displacement, velocity and, in many cases, the acceleration time-histories of recorded earthquakes, and resulted in the following pulse-period--moment-magnitude relation:
\begin{equation}\label{eq:Eq11}
\log_{\text{10}}T_p = -\text{2.2}+\text{0.4}M_\text{W}
\end{equation}
By further accepting self-similar scaling laws between the pulse characteristics and the dimensions of the fault \cite{MavroeidisPapageorgiou2003} proposed:
\begin{equation}\label{eq:Eq12}
\log_{\text{10}}T_p = -\text{2.9}+\text{0.5}M_\text{W}
\end{equation}
and they observed that for the same $M_\text{W}$ the pulse period is larger on average for strike-slip faults than for reverse faults.

\citet{FuMenun2004} after using a similar; yet less sophisticated mathematical expression than Eq. \eqref{eq:Eq10} for the velocity pulse, matched synthetic velocity time-histories generated by the Haskell source model \citep{AkiRichards1980}, and proposed the following velocity pulse period $T_p$ -- moment magnitude $M_\text{W}$ relation:
\begin{equation}\label{eq:Eq13}
\log_{\text{10}}T_p = -\text{3.38}+\text{0.54}M_\text{W}
\end{equation}

\begin{figure}[b!]
\centering
 \includegraphics[width=.95\linewidth]{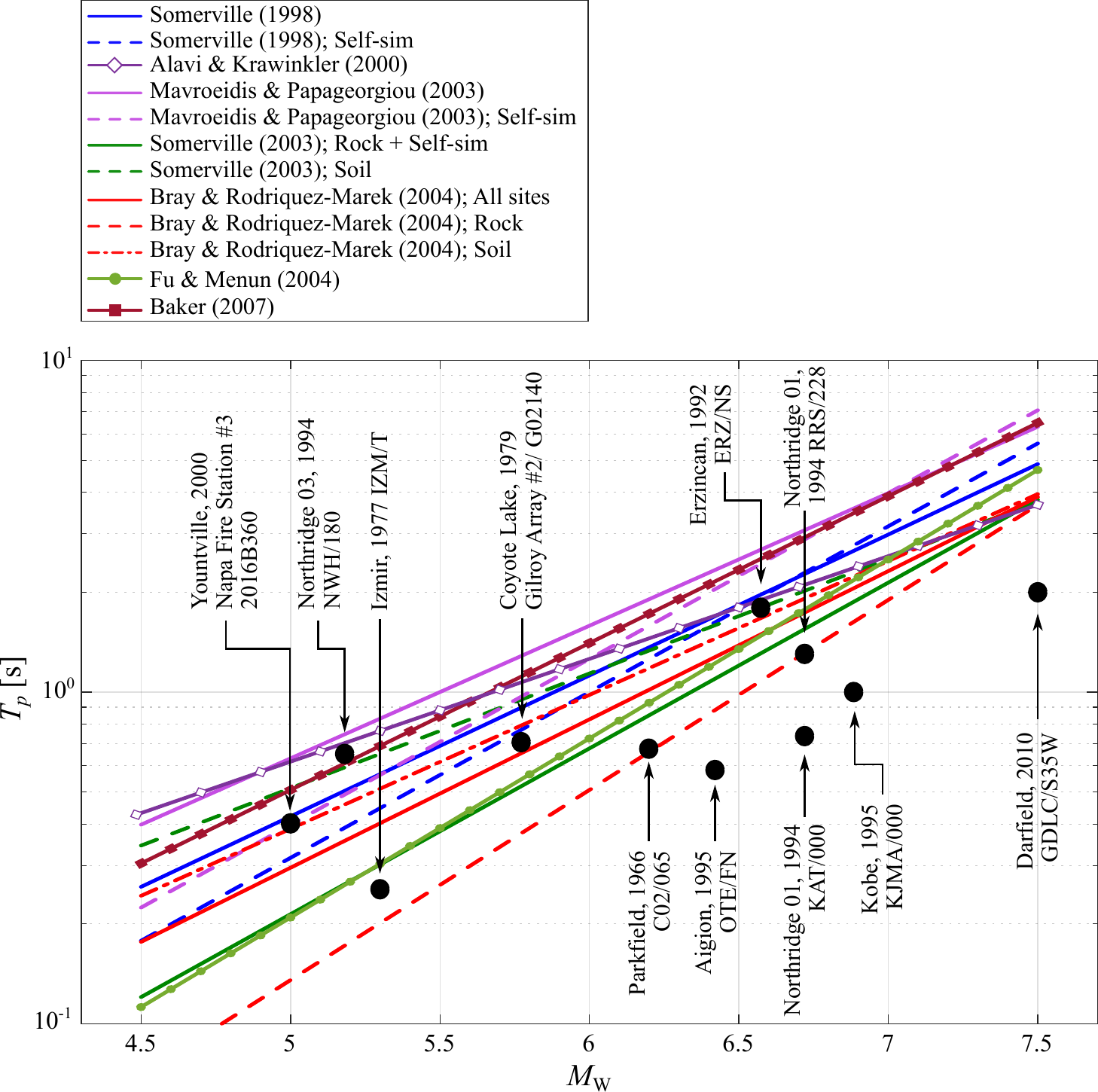}
\caption{Summary of $T_p$--$M_\text{W}$ relationships known to the literature (Eqs. \eqref{eq:Eq04} -- \eqref{eq:Eq14}) that have been proposed by matching invariably velocity pulses together with the pulse periods of selected historic pulse-like records extracted with wavelet analysis on the acceleration records \citep{VassiliouMakris2011}.}
\label{fig:Fig04}
\end{figure}

While the aforementioned studies proposed mathematical expressions of velocity pulses of which the parameters are estimated based on engineering judgement rather than by some formal mathematically objective procedure; the first systematic study for identifying quantitatively coherent velocity pulses in near-fault ground motions was presented by \cite{Baker2007}. Baker's work also focuses on velocity pulses and uses wavelet analysis to automatically extract the largest velocity pulse in a given earthquake record. \cite{Baker2007} employed 4$^\text{th}$-order Daubechies wavelets to distil characteristic time- and length-scales from the recorded ground motions. The main limitation of processing velocity (rather than acceleration) time-histories with wavelet analysis is that one can only extract the visible main velocity pulse --- that is usually the one associated with the near-source effects. \cite{Baker2007} examined the fault-normal components of 398 earthquakes in the Next Generation Attenuation (NGA) database with $M_\text{W} \geq$ 5.5 and recorded within 30 km from the fault. He identified 91 pulse-like records; and through regression analysis of the periods of the identified pulses against the corresponding earthquake moment magnitude, proposed the following pulse-period -- moment-magnitude relation:
\begin{equation}\label{eq:Eq14}
E[\ln T_p] =- \text{5.78} +\text{1.02}M_\text{W}
\end{equation}
where $E[\cdot]$ denotes the expected value of $\ln T_p$. The standard deviation of $\ln T_p$ computed by \cite{Baker2007} is $\sigma_{\ln T_p}=$ 0.55. 

\begin{figure}[t!]
\centering
  \includegraphics[width=.75\linewidth]{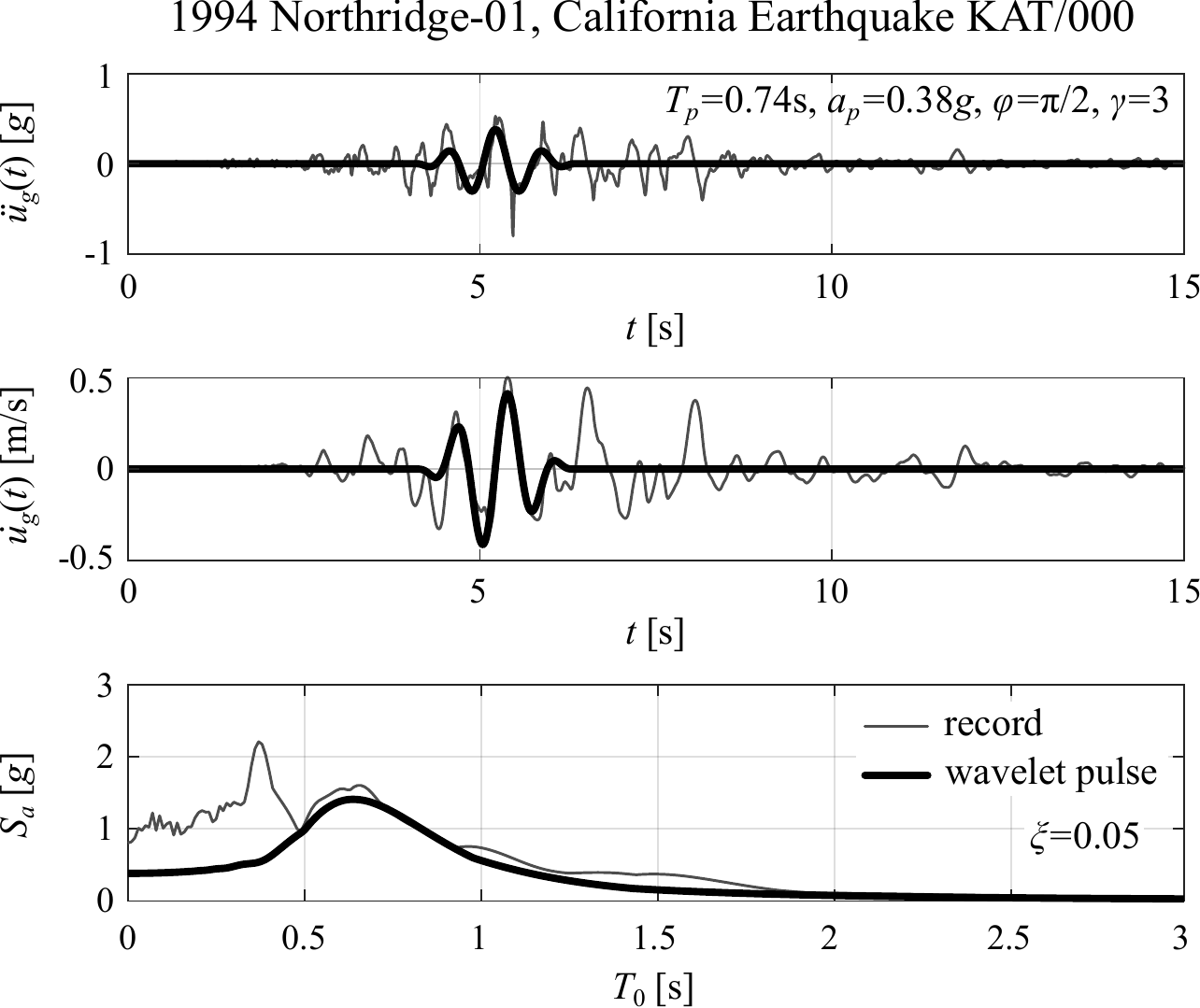}
\caption{Acceleration and velocity time histories together with the associated elastic acceleration spectra of the KAT/000 record from the 1994 Northridge, California earthquake (thin solid lines) and of the M\&P wavelet pulse that best fits the acceleration record (heavy solid lines \citep{VassiliouMakris2011}).}
\label{fig:Fig05}
\end{figure}

Figure \ref{fig:Fig04} plots the 12 pulse-period -- moment-magnitude relations summarized in this section that are all derived invariably by matching velocity pulses of pulse-like ground motions. Clearly, the relations derived for motions recorded on rock offer pulse periods  appreciably lower than the other relations; yet, it is worth recognizing that the differences in the values of the pulse periods offered by the relations plotted in Fig. \ref{fig:Fig04} are large. For instance, for an earthquake of magnitude $M_\text{W}=$ 6.0 the predominant pulse periods range from $T_p \approx$ 0.5 s to $T_p\approx$ 1.6 s. Given that structural deformations (either elastic or inelastic) scale with $T_p^\2$ (as shown by Eqs. \eqref{eq:Eq01}, \eqref{eq:Eq03} and Fig. \ref{fig:Fig03}); when the structural deformations are estimated with $T_p =$ 0.5 s, then $u_\text{max}\sim\text{0.5}^\2=\text{0.25}$; whereas, when  structural deformations are estimated with $T_p =$ 1.6 s, then $u_\text{max}\sim\text{1.6}^\2=\text{2.56}$ --- that is more than a 10-fold increase. This one-order of magnitude variability in the estimation of structural displacements that scale with $a_pT_p^\2$ is the main motivation of this work.

Figure \ref{fig:Fig05} plots the acceleration and velocity time histories together with the elastic acceleration spectra of the KAT/000 ground motion recorded during the 1994 Northridge, California $M_\text{W}=$ 6.7 earthquake (thin solid lines), and of the M\&P wavelet with $T_p=$ 0.74 s, $a_p=$ 0.38$g$, $\phi=\pi/\text{2}$ and $\gamma=$ 3 (heavy solid lines) which is a sister wavelet of the best matching wavelet of the C02 record shown in Fig. \ref{fig:Fig02}.

\section{On the Engineering Significance of Fitting Velocity Pulses}\label{sec:Sec03}
\vspace{-0.5cm}

\begin{figure}[t!]
\centering
  \includegraphics[width=.7\linewidth]{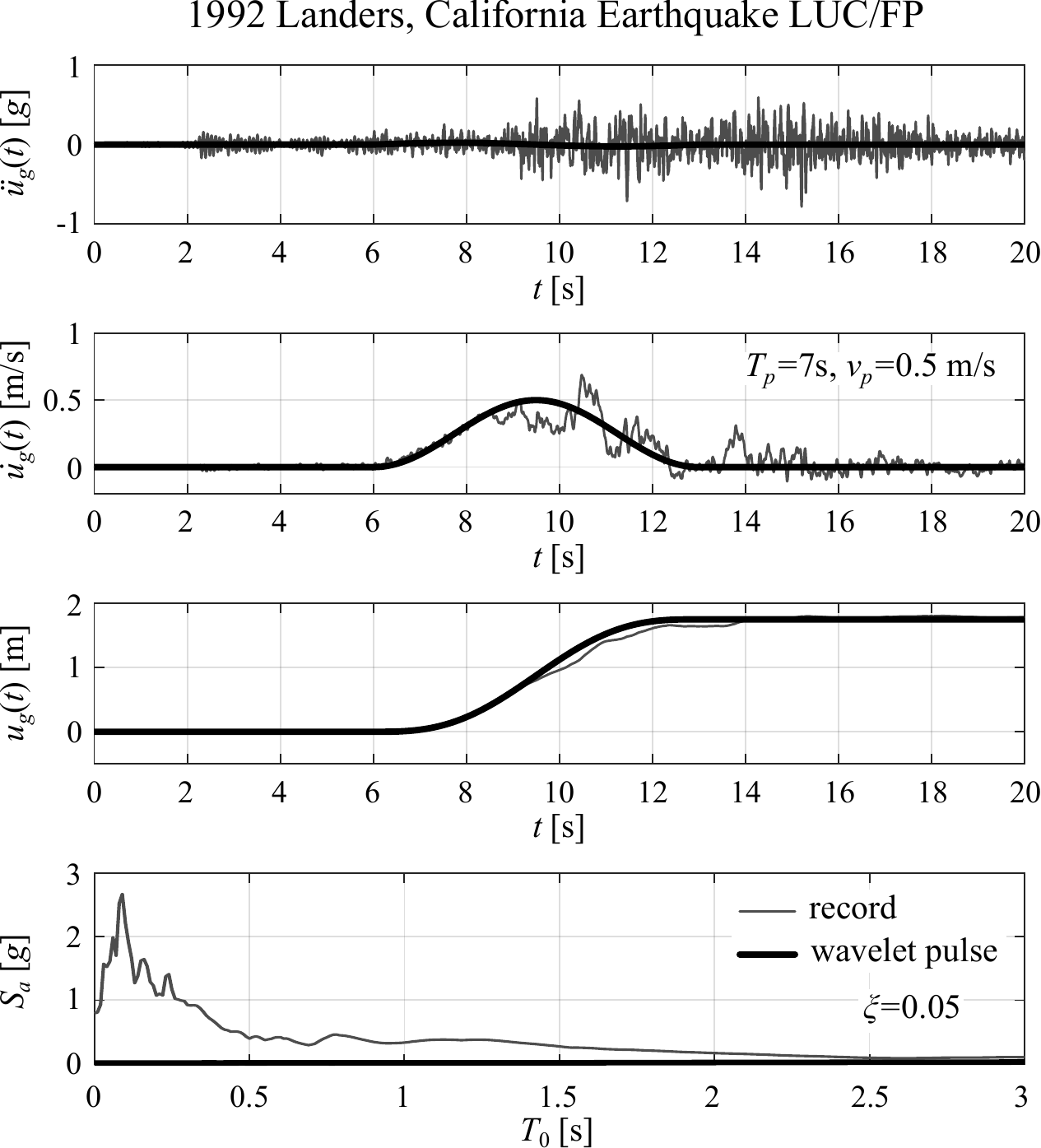}
\caption{Fault parallel component of the acceleration, velocity and displacement histories together with the elastic acceleration response spectra of the June 18, 1992 Landers, California earthquake ($M_\text{W}=$ 7.2) recorded at the Lucerne Valley station (thin solid lines) and a one-sine acceleration pulse that captures the long velocity pulse (heavy solid lines).} 
\label{fig:Fig06}
\end{figure}

\begin{figure}[t!]
\centering
  \includegraphics[width=.7\linewidth]{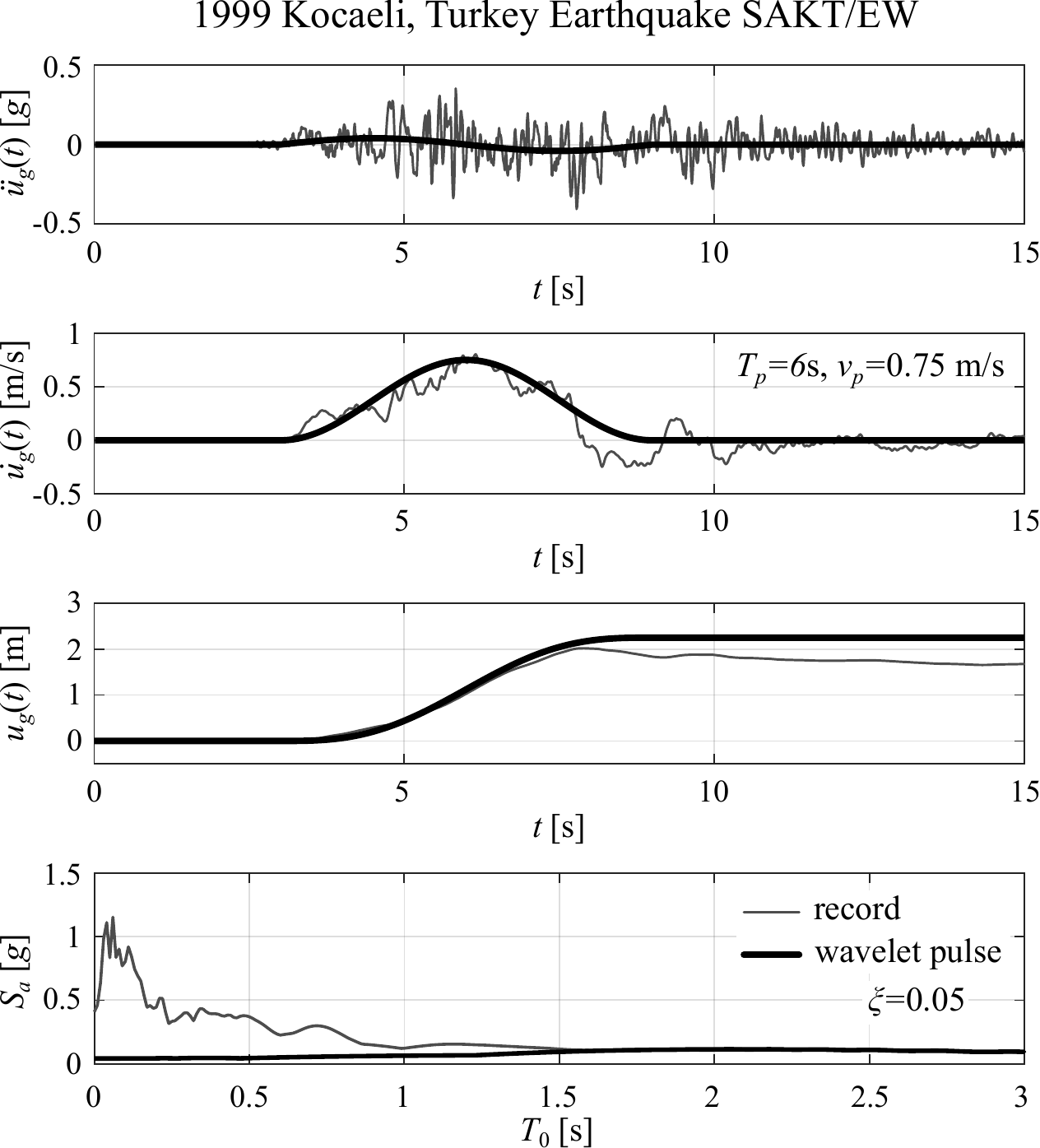}
\caption{East--West component of the acceleration, velocity and displacement histories together with the elastic acceleration response spectra of the August 17, 1999 Kocaeli, Turkey earthquake ($M_\text{W}=$ 7.4) recorded at the Sakarya station and a one-sine acceleration pulse that captures the long velocity pulse (heavy solid lines).}
\label{fig:Fig07}
\end{figure}

Figures \ref{fig:Fig01}, \ref{fig:Fig02} and \ref{fig:Fig05} show examples of ground motions where the velocity pulse is also distinguishable in the acceleration history, and such ground motions are particularly destructive to the built environment. In other cases, acceleration records contain high-frequency spikes and resemble random motions; however, their velocity and displacements histories uncover a coherent long-duration pulse that results from the non-zero mean of the acceleration fluctuations. As an example, Fig. \ref{fig:Fig06} shows the fault parallel components of the acceleration, velocity and displacement histories of the June 18, 1992 Landers, California earthquake ($M_\text{W}=$ 7.2) recorded at the Lucerne Valley station \citep{IwanChen1994}. The coherent 7-s long pulse, responsible for the 1.8-m forward displacement, can also be distinguished in the velocity history; whereas the acceleration history is crowded with high-frequency spikes without exhibiting any visible acceleration pulse. The large spectral accelerations in the period range of $T_\0\approx$ 0.25 s are due to the high-frequency acceleration spikes; whereas the 7-s long velocity pulse that creates the ``biblical'' 1.8-m forward displacement is immaterial to any structure built away from the rupture. A similar situation is shown in Fig. \ref{fig:Fig07} which plots the East--West components of the acceleration, velocity and displacement histories of the August 17, 1999 Kocaeli, Turkey earthquake ($M_\text{W}=$ 7.4) recorded at the Sakarya station. A coherent 6-s long pulse, responsible for the 2.0-m forward displacement, dominates the velocity time history; however, the acceleration history is crowded with high-frequency spikes without revealing any visible pulse. The large spectra accelerations shown at the bottom of Fig. \ref{fig:Fig07} in the range of $T_\0=$ 0.25 s are due to the high-frequency acceleration spikes.

\begin{figure}[t!]
\centering
  \includegraphics[width=\linewidth]{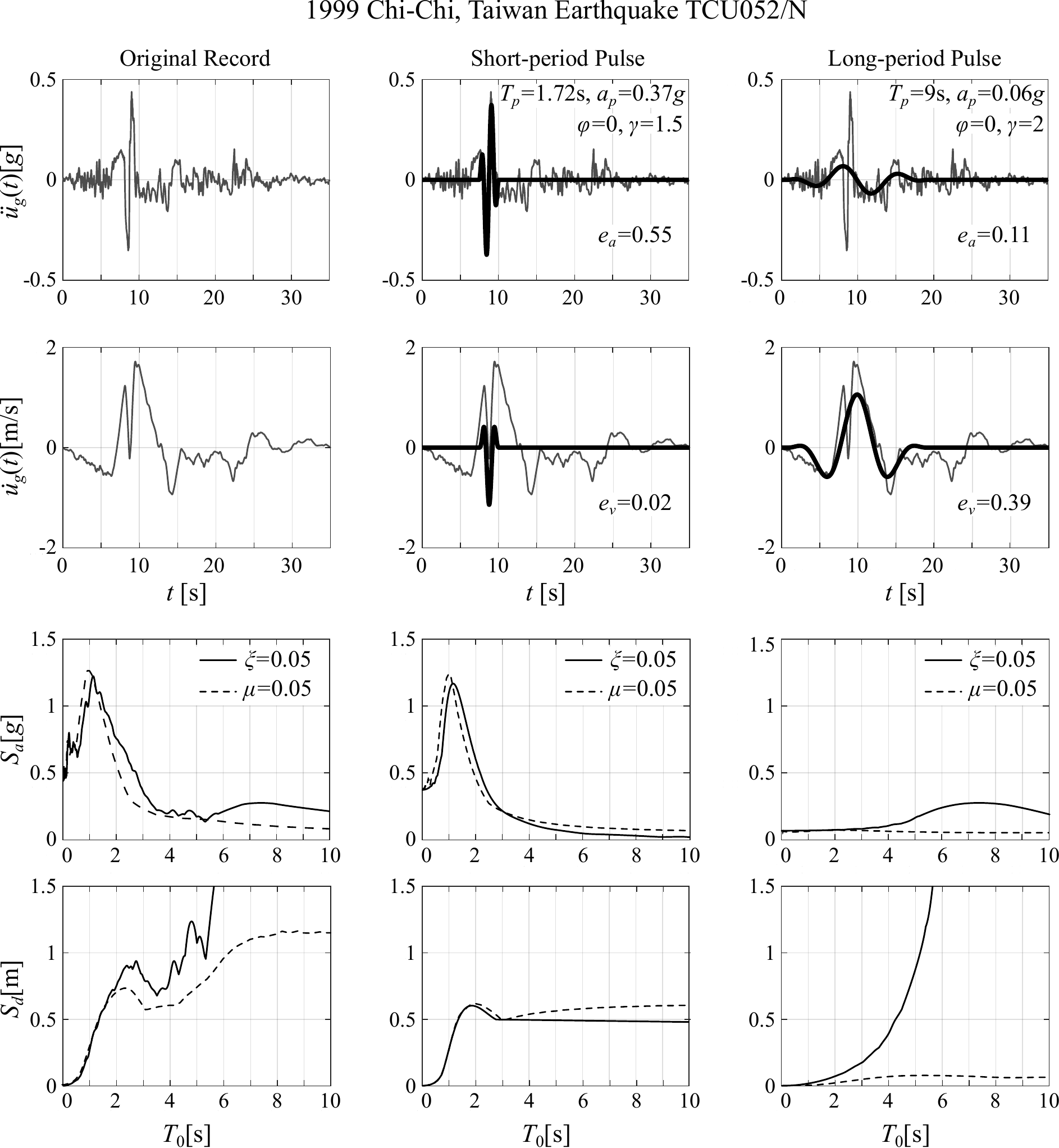}
\caption{Top: Acceleration, velocity and displacement time histories of the TCU052 North--South component from the 1999 Chi-Chi, Taiwan earthquake. Heavy lines denote the short- and long-duration pulses obtained with wavelet analysis. Bottom: Acceleration and displacement spectra of a linear ($m$; $k$; $c$) and a frictional ($m$; $k$; $\mu$) oscillator subjected to the recorded motion (left), the short-period pulse (center), and the long-period pulse that is associated with the near-source effects (right).}
\label{fig:Fig08}
\end{figure}

\cite{Baker2007} recognized that in several occasions of his velocity-centric analysis (in which all daughter wavelets had the same energy --- the default setting in \cite{MATLAB2017} toolbox) the pulse periods obtained from the wavelet analysis differ significantly from the pulse periods obtained from spectral analysis (see Fig. 12 of his paper), and he correctly explained that the pulse period extracted from the peak spectral values is associated in general with the high-frequency oscillatory component of the ground motion; whereas the pulse extracted with the wavelet analysis is associated with the long velocity pulse.

Another situation that challenges the engineering significance of classifying records by only interrogating velocity pulses is that in several near-source records, in addition to the coherent long-period pulse associated with the near-source effect, there is a shorter duration distinguishable pulse (not necessarily of random character) that overrides the larger duration pulse. These shorter duration overriding pulses may be of significant engineering interest to a wide family of structures and there is a clear need to identify and characterize them. For instance, Fig. \ref{fig:Fig08} shows the North--South component of the acceleration, velocity and displacement histories recorded at TCU052 station during the September 21, 1999 Chi-Chi, Taiwan earthquake. This record contains a 9-s long velocity pulse (the one associated with the near-source effects), which is disturbed by a shorter, distinguishable pulse of duration about 1.7 s. This shorter, overriding pulse is of major engineering significance because it is responsible for most of the base-shear and peak deformations of the majority of structures that are of interest in civil engineering as shown in the elastic and inelastic response spectra shown at the bottom of Fig. \ref{fig:Fig08}. Consequently, in order to estimate meaningful structural deformations and base shears it becomes evident that there is a need to extract in a mathematically objective way significant acceleration pulses (not velocity pulses) and construct revised pulse-period -- moment-magnitude relations that emerge by best matching acceleration pulses.

\section{Extraction of Acceleration Pulses with Wavelet Analysis}\label{sec:Sec04}
\vspace{-0.5cm}

\begin{figure}[b!]
\centering
  \includegraphics[width=\linewidth]{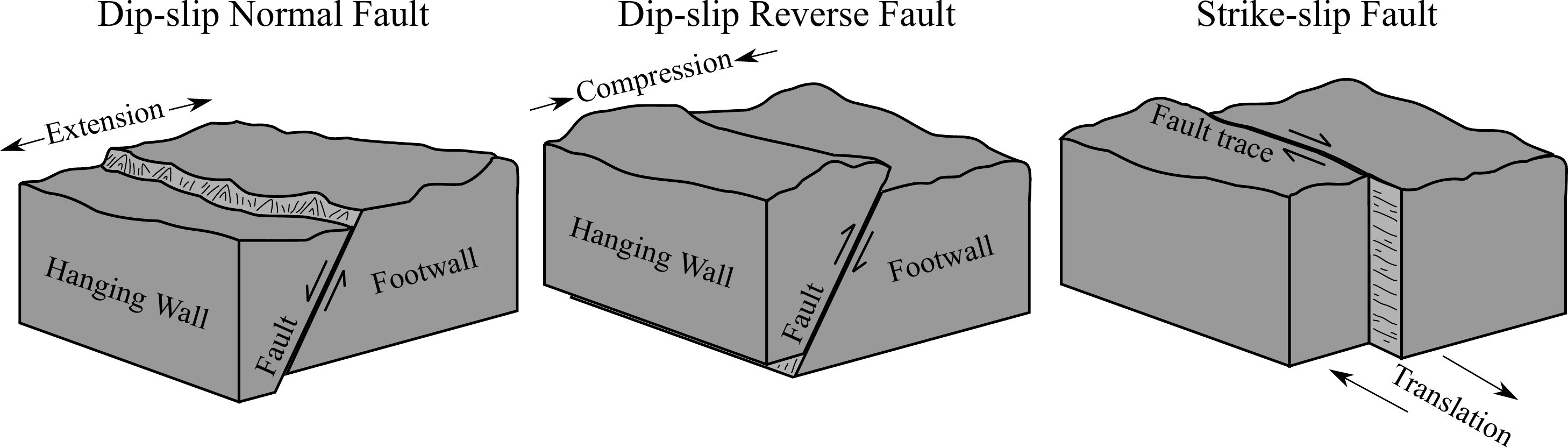}
\caption{Schematic representation of normal reverse, and strike-slip faults. Normal faults are the result of extension while reverse faults are due to compression. Strike-slip faults are the result of translation between the two fault planes.}
\label{fig:Fig09}
\end{figure}

In addition to the pressing need to have estimates of pulse period, $T_p$ associated with acceleration pulses, our paper is partly motivated from the observations of \cite{Somerville1998} and \cite{MavroeidisPapageorgiou2003} who reported that for the same moment magnitude $M_\text{W}$ the pulse periods of ground motions generated from strike-slip faults are on average larger than these from dip-slip reverse faults. Accordingly, a total of 1260 strong-motions with distances from the fault less than 20 km were categorized based on the three faulting mechanisms shown in Fig. \ref{fig:Fig09} --- that is a dip-slip normal fault with a downward moving hanging wall, a dip-slip reverse fault with an upward moving hanging wall and strike-slip fault. Among the 1260 strong-motion records, 447 records are generated by normal faults, 478 records are from reverse faults and and 335 records are generated from strike-slip faults. From the 1260 records, 748 records are available in the Next Generation Attenuation database \citep{AnchetaDarraghStewartSeyhanSilvaChiouWooddellGravesKottkeBooreKishidaDonahue2013} and the remaining 512 records are available in the European Strong Motion Database \citep{AmbraseysSmitSigbjornssonSuhadolcMargaris2002} (see Data and Resources Section).

Forward directivity is observed when the fault rupture propagates towards the site with rupture velocity approximately equal to the shear wave velocity and hence, a coherent and long-period pulse is observed at the beginning of the recorded motion which contains most of the elastic energy. Permanent translation, on the other hand, comes from permanent fault translation as a result of earthquake movement. Previous research studies, such as \citet{Somerville1998, Somerville2003, MavroeidisPapageorgiou2003, BrayRodriguezMarek2004}, focus only on records that presented forward directivity. On the other hand, \citet{Baker2007} did not focus on specific fault mechanisms; yet, acknowledged that a number of the pulse-like earthquake that he identified by  wavelet analysis of his database were the result of forward directivity. Following \citet{Baker2007}, forward directivity effects have not been used as a criterion to choose the near-fault ground motions included in our database. 

The extracted acceleration pulses with wavelet analysis and the resulting velocity pulses have the same effects on the structural response regardless of directivity effects being considered. Wavelets are simple wavelike functions localized on the time axis. For instance, the second derivative of the Gaussian distribution, $e^{-\nicefrac{t^\2}{\2}}$, known in the seismological literature as the symmetric Ricker wavelet \citep{Ricker1943,Ricker1944} and widely referred to as the Mexican Hat wavelet \citep{Addison2002, Addison2017}
\begin{equation}\label{eq:Eq15}
\psi(t)=(\1-t^\2)e^{-\nicefrac{t^\2}{\2}}
\end{equation}
is a wavelike function that can be classified as a wavelet. In  order for a wavelike function to be classified as a wavelet, the wavelike function must have (1) finite energy
\begin{equation}\label{eq:Eq16}
E=\int_{-\infty}^\infty \left| \psi(t)\right| ^\2\, \mathrm{d}t < \infty
\end{equation}
and (2) a zero mean.

Our effort concentrates on achieving the best local matching of any given acceleration record with a wavelet that will offer the best estimates of the period ($T_p=$ time scale) and amplitude ($a_p$, because $a_pT_p^\2=$ length scale)
of the prevailing energetic pulse. Accordingly, we perform a series of inner products (convolutions) of the ground acceleration signal, $\ddot{u}_g(t)$,with the wavelet $\psi(t)$ by manipulating the wavelet through a process of translation (i.e., movement along the time axis) and a process of dilation--contraction (i.e. spreading out or squeezing of the wavelet):
\begin{equation}\label{eq:Eq17}
C(s, \xi)=w(s)\int_\text{0}^\infty \ddot{u}_g(t) \psi(t)\left( \frac{t-\xi}{s} \right)\, \mathrm{d}t
\end{equation}
The values of $s=S$ and $\xi=\Xi$ for which the coefficient, $C(s, \xi)=C(S, \Xi)$, reaches the maximum value offer the scale and location of the wavelet $w(s)\psi\left( \dfrac{t-\xi}{s} \right)$ that locally best matches the acceleration record, $\ddot{u}_g(t)$. Equation \eqref{eq:Eq17} is the definition of the wavelet transform \citep{Addison2017}. The quantity $w(s)$ outside the integral in Eq. \eqref{eq:Eq17} is a weighting function. Typically, $w(s)$ is set equal to $\dfrac{\1}{\sqrt{s}}$ in order to ensure that all wavelets $\psi_{s\comma\,\xi}(t)=w(s)\psi\left( \dfrac{t-\xi}{s} \right)$ at every scale, $s$, have the same energy, and according to Eq. \eqref{eq:Eq16},
\begin{equation}\label{eq:Eq18}
\int_\text{0}^\infty \left| \psi_{s\comma\,\xi}(t) \right|^\2\, \mathrm{d}t = \int_\text{0}^\infty \left| \frac{\1}{\sqrt{s}} \psi\left( \dfrac{t-\xi}{s} \right) \right|^\2\, \mathrm{d}t = \left| \left| \psi_{s\comma\,\xi}(t)   \right| \right|_\2 = \text{ constant} \enskip \forall  \enskip s.
\end{equation}
The same energy requirement among all the daughter wavelets $\psi_{s\comma\,\xi}(t)$ is the default setting in the \cite{MATLAB2017} Wavelet Toolbox; however, the same energy requirement is, by all means, not a restriction. Clearly, there are applications where it is more appropriate that all daughter wavelets, $\psi_{s\comma\,\xi}(t)$, at every scale, $s$, enclose the same area (not same energy) and, in this case, $w(s)=\dfrac{\1}{s}$; therefore \citep{VassiliouMakris2011},
\begin{equation}\label{eq:Eq19}
\int_\text{0}^\infty \left| \psi_{s\comma\,\xi}(t) \right|\, \mathrm{d}t = \int_\text{0}^\infty \left| \frac{\1}{s} \psi\left( \dfrac{t-\xi}{s} \right) \right|\, \mathrm{d}t = \left| \left| \psi_{s\comma\,\xi}(t)   \right| \right|_\1 = \text{ constant} \enskip \forall  \enskip s.
\end{equation}
On the other hand, there may be applications where it is more appropriate that all daughter wavelets have merely the same maximum value and, in this case, $w(s)=$ 1 and
\begin{equation}\label{eq:Eq20}
\left| \left| \psi_{s\comma\,\xi}(t)   \right| \right|_\infty = \text{ constant} \enskip \forall  \enskip s.
\end{equation}

The need to include four parameters in a mathematical expression of a simple wavelike function that is a good candidate to express the coherent component of a recorded ground motion was first recognized and addressed by \cite{MavroeidisPapageorgiou2003} with the velocity pulse expressed by Eq. \eqref{eq:Eq10}. Clearly the wavelike signal given by Eq. \eqref{eq:Eq10} does not always have a zero mean; therefore, it cannot be a wavelet within the context of wavelet transform. Nevertheless, the time derivative of the elementary velocity signal given by Eq. \eqref{eq:Eq10} after setting $t_\0=$ 0 is:
\begin{align}\label{eq:Eq21}
 \frac{\mathrm{d}v(t)}{\mathrm{d}t} & = - \dfrac{\pi f_p}{\gamma} \left\lbrace \sin \left( \dfrac{\text{2}\pi f_p}{\gamma} t\right)\cos ( \text{2}\pi f_p t+\phi) + \gamma \sin ( \text{2}\pi f_p t+\phi)\left[ \text{1} + \cos \left( \dfrac{\text{2}\pi f_p}{\gamma} t \right) \right] \right\rbrace \text{,} \\
  & \quad \quad \quad \quad \quad \quad \quad \quad \quad \quad \quad \quad \quad \quad \quad \quad \quad \quad \quad \quad \quad \quad \quad \quad - \dfrac{\gamma}{\text{2}f_p} \leq  t \leq   \dfrac{\gamma}{\text{2}f_p}  \nonumber
\end{align}
which is, by construction, a zero-mean signal; and was coined by \cite{VassiliouMakris2011} as the \textit{Mavroeidis and Papageorgiou (M\&P) wavelet}. After replacing the oscillatory frequency, $f_p$, with the inverse of the scale parameter, $\dfrac{\1}{s}$ the M\&P wavelet expressed in Eq. \eqref{eq:Eq21} is defined as
\begin{align}\label{eq:Eq22}
\psi\left( \dfrac{t-\xi}{s}\text{, } \gamma \text{, } \phi \right) & = \left\lbrace \sin \left[ \dfrac{\text{2}\pi}{s\gamma} (t-\xi)\right]\cos \left[ \dfrac{\text{2}\pi}{s} (t-\xi) + \phi \right]\right\rbrace \\
& + \gamma \sin \left[ \dfrac{\text{2}\pi}{s} (t-\xi)+\phi\right] \left\lbrace \text{1} + \cos \left[ \dfrac{\text{2}\pi}{s\gamma} (t-\xi)\right] \right\rbrace \text{,}  \quad \xi - \dfrac{\gamma}{\text{2}f_p} \leq  t \leq \xi + \dfrac{\gamma}{\text{2}f_p}  \nonumber
\end{align}
after removing the multiplication factor, $\dfrac{\pi f_p}{\gamma}$, appearing in Eq. \eqref{eq:Eq21}. In the expression for the M\&P wavelet given by Eq. \eqref{eq:Eq22}, the dilation-contraction is controlled with the
parameter $s$ while the movement of the wavelet along the time axis is controlled with translation parameter $\xi$, the same way as is done in the Ricker wavelet given by Eq. \eqref{eq:Eq15}. The novel attraction in the M\&P wavelet given by Eq. \eqref{eq:Eq22} is that, in addition to the dilation-contraction and translation $\dfrac{t-\xi}{s}$, the wavelet can be further manipulated by modulating the phase, $\phi$, and the parameter, $\gamma$, which controls the oscillatory character (number of half cycles). Accordingly, \cite{VassiliouMakris2011} proposed the four-parameter wavelet transform as:
\begin{equation}\label{eq:Eq23}
C(s\comma\, \xi\comma\, \gamma\comma\, \phi)=w(s\comma\, \gamma\comma\, \phi) \int_\text{0}^\infty \ddot{u}_g(t)\psi\left( \dfrac{t-\xi}{s}\comma\, \gamma\comma\, \phi \right) \, \mathrm{d}t.
\end{equation}
The inner product, given by Eq. \eqref{eq:Eq23}, is performed repeatedly by scanning not only at all times, $\xi$, and all scales, $s$, but also by scanning various phases, $\phi=\left\lbrace \0\comma\, \pi/\4\comma\, \pi/\2\comma\,\3\pi/\4\comma\,\2\pi\right\rbrace$, and various values of the oscillatory nature of the signal $\gamma=\left\lbrace \1\comma\, \text{1.5}\comma\, \2\comma\, \text{2.5}\comma\, \3 \right\rbrace$. When needed, more values of $\phi$ and $\gamma$ may be scanned. The quantity $w(s\comma\, \gamma\comma\, \phi) $ outside the integral is a weighting function, which is adjusted according to the application. The values of $s=S$, $\xi=\Xi$, $\gamma=\Gamma$, $\phi=\Phi$, for which the coefficient $C(s\comma\, \xi\comma\, \gamma\comma\, \phi)=C(S\comma\, \Xi\comma\, \Gamma\comma\, \Phi)$ reaches its maximum value, offer the scale, time location, phase, and number of half cycles of the wavelet $\psi\left( \dfrac{t-\xi}{s}\comma\, \gamma\comma\, \phi \right)$ that locally best matches the acceleration record, $\ddot{u}_g(t)$. The multiplication coefficient, $\lambda(S\comma\, \Xi\comma\, \Gamma\comma\, \Phi)$, which dictates how much the best-matching generalized wavelet $w(S\comma\, \Gamma\comma\, \Phi)\cdot \psi_{S\comma\, \Xi\comma\, \Gamma\comma\, \Phi}(t)$ needs to be amplified to best approximate the energetic acceleration pulse, is \citep{VassiliouMakris2011}
\begin{equation}\label{eq:Eq24}
\lambda(S\comma\, \Xi\comma\, \Gamma\comma\, \Phi)=\frac{C(S\comma\, \Xi\comma\, \Gamma\comma\, \Phi)}{w^\2 (S\comma\, \Gamma\comma\, \Phi)\cdot S \cdot E(\Gamma\comma\, \Phi)}.
\end{equation}
For instance, the mathematical wavelets that best match the acceleration records shown in Figs. \ref{fig:Fig01} and \ref{fig:Fig02} have been extracted with the extended wavelet transform defined by Eq. \eqref{eq:Eq23}.

\section{Pulse-Period -- Moment-Magnitude Relations with Engineering Significance}\label{sec:Sec05}
\vspace{-0.5cm}
The initial motivation of \cite{BerteroHerreraMahin1976,BerteroMahinHerrera1978} to divert the attention of engineers to coherent acceleration pulses and the ongoing interest in identifying and characterizing them are primarily motivated from the need to estimate inelastic deformation demands on structures like those plotted in Fig. \ref{fig:Fig03}. Clearly, the more a ground motion is pulse-like, the more the results from self-similar master curves like the one presented by \cite{MakrisPsychogios2006} and \cite{KaravasilisMakrisBazeosBeskos2010} become dependable. Accordingly, some indicator is needed to indicate to what extent a recorded ground motion is pulse-like. Efforts to classify ground motions have been presented in the past. \cite{Baker2007}, after a preliminary visual classification of the velocity records, proceeds by proposing a pulse indicator that is a function of a \textit{PGV} ratio and an energy ratio. Clearly, both the \textit{PGV} ratio and the energy ratio involve information solely from the velocity time history, and therefore Baker’s pulse indicator does not have the ability to identify shorter duration pulses that occasionally govern to a great extent the response of structures as shown in Fig. \ref{fig:Fig08}. 

\begin{figure}[htbp!]
  \centering
  \includegraphics[width=.5\linewidth]{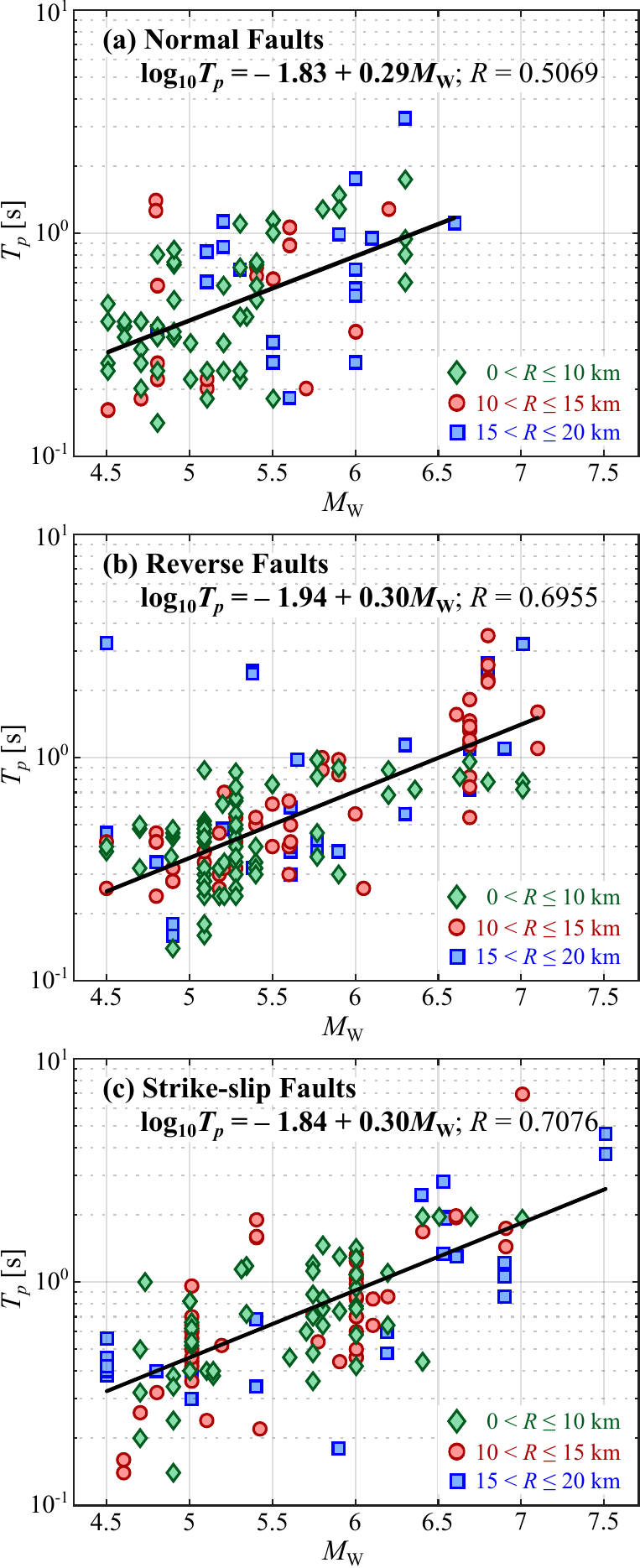}
  \caption{Pulse periods extracted with the extended wavelet transform (Eq. \eqref{eq:Eq23}) of the 109 pulse-like records (PI $\geq$ 0.3) generated from normal faults (a); the 188 pulse-like records (PI $\geq$ 0.3) generated from reverse faults (b); and the 125 pulse-like records (PI $\geq$ 0.3) generated from strike-slip faults (c).}
  \label{fig:Fig10}
\end{figure}

In view of the engineering significance that emerges when a shorter-duration pulse overrides a long-duration pulse as illustrated in Fig. \ref{fig:Fig08}, \cite{VassiliouMakris2011} proposed a pulse indicator (PI)
\begin{equation}\label{eq:Eq25}
\text{PI} = \dfrac{\text{1}}{\text{2}}(e_a + e_v)
\end{equation}
where
\begin{equation}\label{eq:Eq26}
e_a = \dfrac{\displaystyle\int_{\text{0}}^{\infty}\ddot{u}(t) \cdot \lambda(S\comma\, \Xi\comma\, \Gamma\comma\, \Phi)  w(S\comma\, \Gamma\comma\, \Phi)\psi_{S\comma\, \Xi\comma\, \Gamma\comma\, \Phi}(t)        \mathrm{d}t}{\displaystyle\int_{\text{0}}^{\infty}[\ddot{u}(t)]^{\text{2}}\mathrm{d}t} 
\end{equation}
is a scalar measure of the performance of the best matching wavelet, $\lambda(S\comma\, \Xi\comma\, \Gamma\comma\, \Phi) w(S\comma\, \Gamma\comma\, \Phi)$ to locally match the predominant acceleration pulse and
\begin{equation}\label{eq:Eq27}
e_v = \dfrac{\displaystyle\int_{\text{0}}^{\infty}\dot{u}(t)\cdot v(t)\mathrm{d}t}{\displaystyle\int_{\text{0}}^{\infty}[\dot{u}(t)]^{\text{2}}\mathrm{d}t} 
\end{equation}
where $v(t)$ is the velocity pulse associated with the extracted acceleration pulse $\lambda(S\comma\, \Xi\comma\, \Gamma\comma\, \Phi)  \cdot w(S\comma\, \Gamma\comma\, \Phi)\psi_{S\comma\, \Xi\comma\, \Gamma\comma\, \Phi}(t)$
\begin{equation}\label{eq:Eq28}
v(t)=\int \lambda(S\comma\, \Xi\comma\, \Gamma\comma\, \Phi)  w(S\comma\, \Gamma\comma\, \Phi)\psi_{S\comma\, \Xi\comma\, \Gamma\comma\, \Phi}(t) \, \mathrm{d}t
\end{equation}
The matching indices $e_a$ and $e_v$ of all 1260 candidate records with epicentral distance $D\leq$ 20 km were computed and each record that exhibited a pulse indicator PI $=\dfrac{\text{1}}{\text{2}}(e_a + e_v)\geq$ 0.3 was classified as a pulse-like record \cite{VassiliouMakris2011}. Accordingly, among the 447 normal fault records with epicentral distance $D\leq$ 20 km, 109 records were classified as pulse-like (PI $\geq$ 0.3)  with their pulse periods shown in Fig. \ref{fig:Fig10} (top). Similarly, among the 478 reverse fault records with epicentral distance $D\leq$ 20 km, 188 records were classified as pulse-like (PI $\geq$ 0.3) with their pulse periods shown in Fig. \ref{fig:Fig10} (center); whereas, among the 335 strike-slip fault records with epicentral distance $\leq$ 20 km, 125 records were classified as pulse-like (PI $\geq$ 0.3)  with their pulse periods shown in Fig. \ref{fig:Fig10} (bottom). For the complete list of the pulse-like ground motions and the characteristics of the extracted pulses, see Tables \ref{tab:TabS1}, \ref{tab:TabS2} and \ref{tab:TabS3} in the supplement at the end of this article.

The extracted pulse periods with the extended wavelet transform given by Eq. \eqref{eq:Eq23} of the pulse-like records with epicentral distance $D\leq$ 10 km are shown in Fig. \ref{fig:Fig10} with diamonds. Similarly, the extracted pulse periods of the pulse-like records with epicentral distance 10 km $<D\leq$ 15 km are shown in Fig. \ref{fig:Fig10} with open circles; whereas, the extracted pulse periods of the pulse-like records with epicentral distance 15 km $<D\leq$ 20 km are shown with open squares.

The straight solid lines in Fig. \ref{fig:Fig10} result from regression analysis of all the extracted data ($D\leq$ 20 km) appearing in each subplot and are:
\begin{equation}\label{eq:Eq29}
\log_{\text{10}}T_{p_{\text{ NF}}} = -\text{1.83}+\text{0.29}M_W \text{, \quad with } \quad R=\text{0.5069}
\end{equation}
for dip-slip normal faults (NF),
\begin{equation}\label{eq:Eq30}
\log_{\text{10}}T_{p_{\text{ RF}}} = -\text{1.94}+\text{0.30}M_W \text{, \quad with } \quad R=\text{0.6955}
\end{equation}
for dip-slip reverse faults (RF), and
\begin{equation}\label{eq:Eq31}
\log_{\text{10}}T_{p_{\text{ SSF}}} = -\text{1.84}+\text{0.30}M_W \text{, \quad with } \quad R=\text{0.7076}
\end{equation}
for strike-slip faults (SSF). In Eqs. \eqref{eq:Eq29} -- \eqref{eq:Eq31}, $R$ is the correlation coefficient \citep{Weisberg2005}. 

\begin{figure}[t!]
\centering
  \includegraphics[width=.95\linewidth]{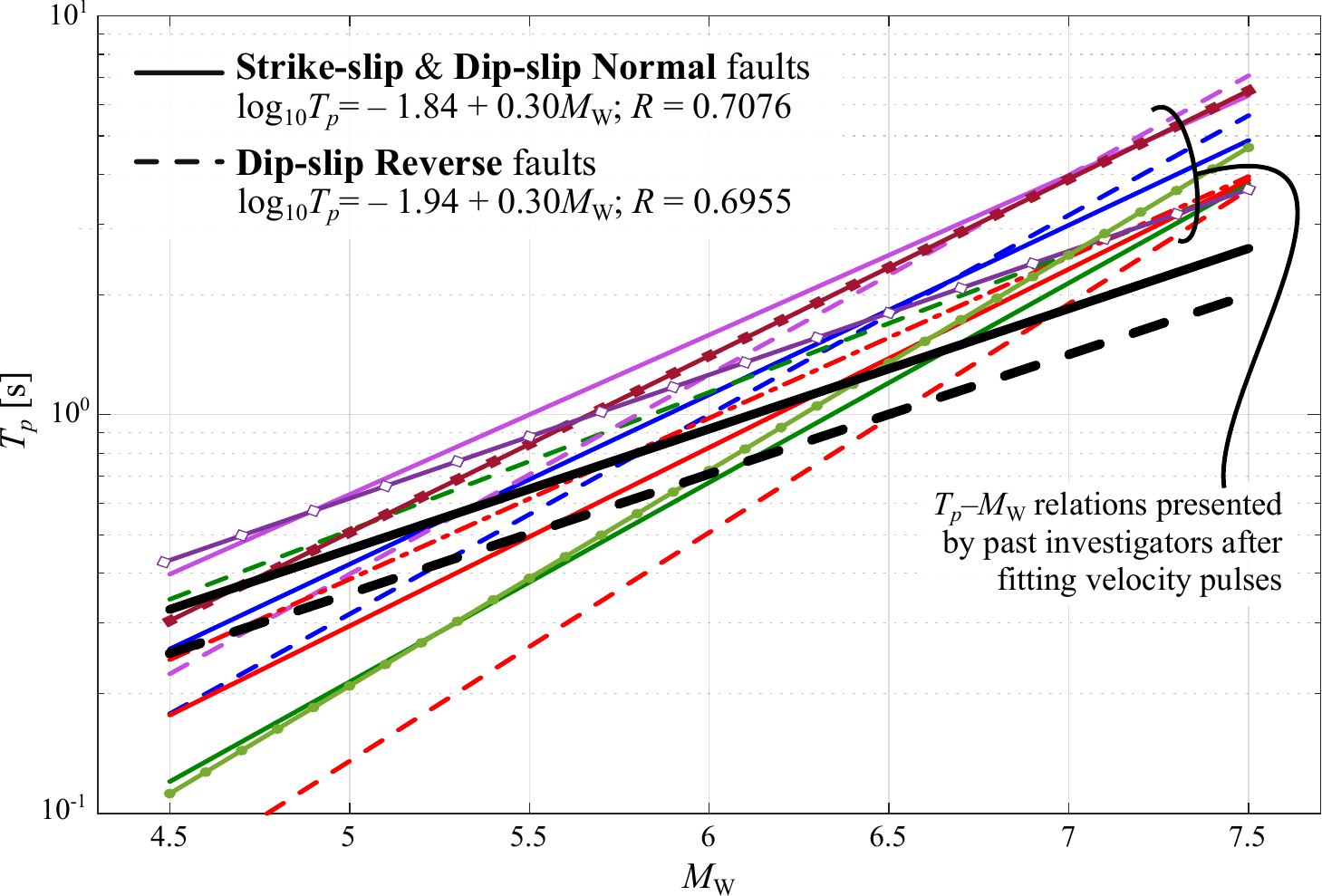}
\caption{Proposed $T_p$--$M_\text{W}$ relations for pulse-like ground motions with epicentral distance $D\leq$ 20 km for strike-slip and normal dip-slip faults (heavy solid line: Eq. \eqref{eq:Eq31}) and for reverse dip-slip faults (heavy dashed line: Eq. \eqref{eq:Eq30}) extracted with wavelet analysis on the acceleration records.}
\label{fig:Fig11}
\end{figure}

The proximity of the coefficients of Eqs. \eqref{eq:Eq29} and \eqref{eq:Eq31} suggests that the same $T_p$--$M_\text{W}$ equation can be used for normal dip-slip faults and strike-slip faults; whereas, Eq. \eqref{eq:Eq30} is recommended for reverse dip-slip faults. Figure \ref{fig:Fig11} plots Eqs. \eqref{eq:Eq30} and \eqref{eq:Eq31} that have been derived after interrogating with wavelet analysis acceleration pulse-like records against the published $T_p$--$M_\text{W}$ relations presented in Fig. \ref{fig:Fig04} which have been derived by merely matching velocity pulses. Our findings, that emerge from wavelet analysis on acceleration records, confirm past observations by \cite{Somerville1998} and \cite{MavroeidisPapageorgiou2003} that for the same moment magnitude, $M_\text{W}$, the pulse periods of ground motions generated from strike-slip faults are on average larger than the pulse periods generated from reverse dip-slip faults. Nevertheless, our proposed $T_p$--$M_\text{W}$ relations expressed by Eqs. \eqref{eq:Eq30} and \eqref{eq:Eq31} are shown in Fig. \ref{fig:Fig11} with heavy lines manifest a lower slope than the slopes of the $T_p$--$M_\text{W}$ relations plotted by past investigators after matching velocity pulses. As a result, our proposed relations expressed by Eqs. \eqref{eq:Eq30} and \eqref{eq:Eq31} suggest appreciably shorter pulse periods for ground motions generated by earthquakes with moment magnitude $M_\text{W}\geq$ 6.

\begin{figure}[t!]
\centering
  \includegraphics[width=0.55\linewidth]{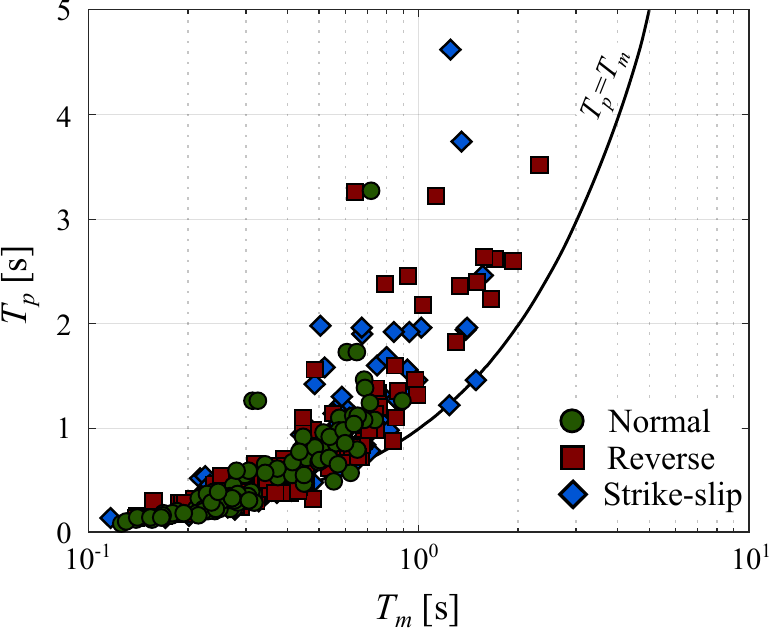}
\caption{Comparison of the pulse period $T_p$ of all the pulse-like ground motions appearing in Fig. \ref{fig:Fig10} which have been extracted with wavelet analysis on the acceleration records together with the period $T_m$ offered by Eq. \eqref{eq:Eq32} \citep{RathjeAbrahamsonBray1998}. }
\label{fig:Fig12}
\end{figure}

The pulse periods of the individual pulse-like ground motions shown in Fig. \ref{fig:Fig10} that have been extracted with wavelet analysis of the acceleration records are compared with the mean period
\begin{equation}\label{eq:Eq32}
T_m = \dfrac{\displaystyle\sum_j C^{\text{2}}_j\tfrac{\text{1}}{f_j}}{\displaystyle\sum_j C^{\text{2}}_j} \quad \text{for 0.25 Hz} \leq f_j \leq \text{20 Hz}
\end{equation}
proposed by \cite{RathjeAbrahamsonBray1998}. In Eq. \eqref{eq:Eq32} $f_j$ are discrete frequencies between 0.25 Hz and 20 Hz; whereas, $C_j$ are the Fourier amplitudes of the Fourier transform of the acceleration record evaluated at frequencies $f_j$. The horizontal axis (logarithmic scale) of Fig. \ref{fig:Fig12} indicates the computed mean period, $T_m$, of all the pulse-like ground motions summarized in Fig. \ref{fig:Fig10}; whereas, the vertical axis (linear scale) indicates the pulse periods of all the pulse-like ground motions appearing in Fig. \ref{fig:Fig10} which have been extracted with wavelet analysis of the acceleration records. Figure \ref{fig:Fig12} uncovers that the mean period, $T_m$, offered by Eq. \eqref{eq:Eq32} and proposed by \cite{RathjeAbrahamsonBray1998} is invariably lower than the pulse periods extracted with wavelet analysis. This is anticipated since Eq. \eqref{eq:Eq32} factors several high-frequency spikes with periods much lower than the period of the dominant distinguishable pulse extracted with wavelet analysis. Accordingly, for earthquakes with $M_\text{W}>$ 6 our proposed $T_p$--$M_\text{W}$ relations expressed by Eqs. \eqref{eq:Eq30} and \eqref{eq:Eq31} offer pulse periods, which are lower than these offered by the $T_p$--$M_\text{W}$ relations proposed by other investigators and summarized in Fig. \ref{fig:Fig04}; yet our pulse periods remain larger than the mean periods, $T_m$, offered by Eq. \eqref{eq:Eq32}.

Consequently, our proposed wavelet analysis on acceleration pulse-like records and the resulting revised $T_p$--$M_\text{W}$ relations yield objective $T_p$ values allowing for the estimation of dependable peak structural displacements that scale invariably with $a_pT_p^\text{2}$.

\section{Conclusions}\label{sec:Sec06}
\vspace{-0.5cm}
We have revisited the pulse-period -- moment-magnitude ($T_p$--$M_\text{W}$) relations of pulse-like ground motions generated from near-source earthquakes with epicentral distances $D\leq$ 20 km. A total of 1260 ground motions are interrogated with wavelet analysis and the associated extended wavelet transform \citep{VassiliouMakris2011} to identify energetic acceleration pulses (not velocity pulses) and extract their optimal period, $T_p$, amplitude $a_p$, phase $\phi$, and number of half-sines $\gamma$.

The interrogation of acceleration records with wavelet analysis is capable of extracting shorter-duration distinguishable pulses with engineering significance which override the longer near-source pulses that are not of random character. The pulse-like character of any given record was established with the implementation of a pulse indicator, PI = $\dfrac{\text{1}}{\text{2}}(e_a+e_v)$ where $e_a$ and $e_v$ are scalar measures of the performance of the best matching wavelet on the acceleration record and the velocity record.

The extended wavelet transform that was employed in this study identified 109 pulse-like records from normal faults, 188 pulse-like records from reverse faults and 125 pulse-like records from strike-slip faults, all with epicentral distances, $D\leq$ 20 km. Regression analysis on the extracted periods concluded that the same $T_p$--$M_\text{W}$  relation can be used for pulse-like ground motions generated either from strike-slip faults or from dip-slip normal faults; whereas, a different $T_p$--$M_\text{W}$ relation is proposed for dip-slip reverse faults.

The study concludes that for the same moment magnitude, $M_\text{W}$, the pulse period of ground motions generated from strike-slip faults are on average larger than these from reverse faults --- a result that is in agreement with findings from past investigators. At the same, our wavelet analysis on acceleration records produces $T_p$--$M_\text{W}$ relations with a lower slope than the slopes of the $T_p$--$M_\text{W}$ relations presented by past investigators who concentrated on velocity pulses. As a result, our proposal $T_p$--$M_\text{W}$ relations yield lower $T_p$ value for larger magnitude earthquakes (say $M_\text{W}>$ 6); yet they remain higher than the mean period, $T_m$, proposed by \cite{RathjeAbrahamsonBray1998}. Accordingly, our results allow for the estimation of dependable peak structural displacements that scale invariably with $a_pT_p^\text{2}$.

\section*{Data and Resources}\label{sec:Data}
\vspace{-0.5cm}
The earthquake records listed in Tables \ref{tab:TabS1}, \ref{tab:TabS2} and \ref{tab:TabS3} in the supplement to this paper were downloaded from:
\begin{enumerate}
\item \url{https://ngawest2.berkeley.edu} (Last accessed April 2021) \vspace*{-0.2cm}
\item \url{http://www.isesd.hi.is} (Last accessed April 2021)
\end{enumerate}

\noindent All the computations, numerical analysis, and plotting have been performed with the software \cite{MATLAB2017}. \url{https://www.mathworks.com} (last accessed April 2021).

\bibliographystyle{myapalike}
\bibliography{References} 

\begin{thebibliography}{}

\bibitem[Addison, 2002]{Addison2002}
Addison, P.~S. (2002).
\newblock {\em The illustrated wavelet transform handbook}.
\newblock Institute of Physics Handbook, London press.

\bibitem[Addison, 2017]{Addison2017}
Addison, P.~S. (2017).
\newblock {\em The illustrated wavelet transform handbook: introductory theory
  and applications in science, engineering, medicine and finance}.
\newblock CRC press.

\bibitem[Aki and Richards, 1980]{AkiRichards1980}
Aki, K. and Richards, P.~G. (1980).
\newblock {\em Quantitative Seismology: Theory and Methods}.
\newblock Freeman.

\bibitem[Alavi and Krawinkler, 2000]{AlaviKrawinkler2000}
Alavi, B. and Krawinkler, H. (2000).
\newblock Consideration of near-fault ground motion effects in seismic design.
\newblock In {\em Proceedings of the 12$^{\text{th}}$ World Conference on
  Earthquake Engineering}, page~8.

\bibitem[Ambraseys \textit{et~al.},
  2002]{AmbraseysSmitSigbjornssonSuhadolcMargaris2002}
Ambraseys, N., Smit, P., Sigbj{\"o}rnsson, R., Suhadolc, P., and Margaris, B.
  (2002).
\newblock Internet-site for {E}uropean strong-motion data.
\newblock {\em European Commission, Research-Directorate General, Environment
  and Climate Programme}.

\bibitem[Ancheta \textit{et~al.},
  2013]{AnchetaDarraghStewartSeyhanSilvaChiouWooddellGravesKottkeBooreKishidaDonahue2013}
Ancheta, T.~D., Darragh, R.~B., Stewart, J.~P., Seyhan, E., Silva, W.~J.,
  Chiou, B.~S., Wooddell, K.~E., Graves, R.~W., Kottke, A.~R., Boore, D.~M.,
  Kishida, T., and Donahue, J.~L. (2013).
\newblock Peer {NGA}-{W}est2 database.
\newblock {\em Pacific Earthquake Engineering Research Center Berkeley, CA}.

\bibitem[Baker, 2007]{Baker2007}
Baker, J.~W. (2007).
\newblock Quantitative classification of near-fault ground motions using
  wavelet analysis.
\newblock {\em Bulletin of the Seismological Society of America},
  97(5):1486--1501.

\bibitem[Barenblatt, 1996]{Barenblatt1996}
Barenblatt, G.~I. (1996).
\newblock {\em Scaling, self-similarity, and intermediate asymptotics:
  dimensional analysis and intermediate asymptotics}.
\newblock Number~14. Cambridge University Press.

\bibitem[Bertero \textit{et~al.}, 1991]{BerteroAndersonKrawinklerMiranda1991}
Bertero, V.~V., Anderson, J.~C., Krawinkler, H., and Miranda, E. (1991).
\newblock Design guidelines for ductility and drift limits.
\newblock {\em Rep. No. UCB/EERC-91}, 15.

\bibitem[Bertero \textit{et~al.}, 1976]{BerteroHerreraMahin1976}
Bertero, V.~V., Herrera, R.~A., and Mahin, S.~A. (1976).
\newblock Establishment of design earthquakes—evaluation of present methods.
\newblock In {\em Proc., Int. Symp. on Earthquake Structural Engineering},
  volume~1, pages 551--580. Univ. of Missouri-Rolla Rolla, Mo.

\bibitem[Bertero \textit{et~al.}, 1978]{BerteroMahinHerrera1978}
Bertero, V.~V., Mahin, S.~A., and Herrera, R.~A. (1978).
\newblock Aseismic design implications of near-fault san fernando earthquake
  records.
\newblock {\em Earthquake Engineering \& Structural Dynamics}, 6(1):31--42.

\bibitem[Bolt, 1971]{Bolt1971}
Bolt, B.~A. (1971).
\newblock The {S}an {F}ernando valley, {C}alifornia, earthquake of {F}ebruary 9
  1971: {D}ata on seismic hazards.
\newblock {\em Bulletin of the Seismological Society of America},
  61(2):501--510.

\bibitem[Bolt, 1975]{Bolt1975}
Bolt, B.~A. (1975).
\newblock How are magnitude, epicenter, focal depth determined? {D}egree of
  accuracy? {D}escribe {P} and {S} waves, etc.
\newblock In {\em Engineering Aspects of the Lima, Peru Earthquake of October
  3, 1974}, pages 74--75. Earthquake Engineering Research Institute.

\bibitem[Bray and Rodriguez-Marek, 2004]{BrayRodriguezMarek2004}
Bray, J.~D. and Rodriguez-Marek, A. (2004).
\newblock Characterization of forward-directivity ground motions in the
  near-fault region.
\newblock {\em Soil dynamics and earthquake engineering}, 24(11):815--828.

\bibitem[FEMA, 2000]{FEMA2000}
FEMA (2000).
\newblock {\em NEHRP Recommended Seismic Provisions for New Buildings and Other
  Structures}.

\bibitem[FEMA-273, 1997]{FEMA273_1997}
FEMA-273 (1997).
\newblock {\em NEHRP Guidelines for the Seismic Rehabilitation of Buildings}.

\bibitem[Fu and Menun, 2004]{FuMenun2004}
Fu, Q. and Menun, C. (2004).
\newblock Seismic-environment-based simulation of near-fault ground motions.
\newblock In {\em Proceedings of the 13$^{\text{th}}$ World Conference on
  Earthquake Engineering}.

\bibitem[Gabor, 1946]{Gabor1946}
Gabor, D. (1946).
\newblock Theory of communication. {P}art 1: {T}he analysis of information.
\newblock {\em Journal of the Institution of Electrical Engineers-Part III:
  Radio and Communication Engineering}, 93(26):429--441.

\bibitem[Hall \textit{et~al.}, 1995]{HallHeatonHallingWald1995}
Hall, J.~F., Heaton, T.~H., Halling, M.~W., and Wald, D.~J. (1995).
\newblock Near-source ground motion and its effects on flexible buildings.
\newblock {\em Earthquake Spectra}, 11(4):569--605.

\bibitem[Housner and Trifunac, 1967]{HousnerTrifunac1967}
Housner, G. and Trifunac, M. (1967).
\newblock Analysis of accelerograms—{P}arkfield earthquake.
\newblock {\em Bulletin of the Seismological Society of America},
  57(6):1193--1220.

\bibitem[Housner and Hudson, 1958]{HousnerHudson1958}
Housner, G.~W. and Hudson, D.~E. (1958).
\newblock The {P}ort {H}ueneme earthquake of {M}arch 18, 1957.
\newblock {\em Bulletin of the Seismological Society of America},
  48(2):163--168.

\bibitem[Housner and Hudson, 1959]{HousnerHudson1959}
Housner, G.~W. and Hudson, D.~E. (1959).
\newblock Applied mechanics-dynamics. vol. ii, princeton, n. j.

\bibitem[IBC, 2000]{IBC2000}
IBC (2000).
\newblock {\em 2000 International Building Code}.

\bibitem[Iwan and Chen, 1994]{IwanChen1994}
Iwan, W. and Chen, X. (1994).
\newblock Important near-field ground motion data from the {L}anders
  earthquake.
\newblock In {\em Proceedings of the 10$^\text{th}$ European Conference on
  Earthquake Engineering}, volume~1, pages 229--234. AA Balkema Rotterdam, The
  Netherlands.

\bibitem[Karavasilis \textit{et~al.}, 2010]{KaravasilisMakrisBazeosBeskos2010}
Karavasilis, T.~L., Makris, N., Bazeos, N., and Beskos, D.~E. (2010).
\newblock Dimensional response analysis of multistory regular steel {MRF}
  subjected to pulselike earthquake ground motions.
\newblock {\em Journal of Structural Engineering}, 136(8):921--932.

\bibitem[Krawinkler and Alavi, 1998]{KrawinklerAlavi1998}
Krawinkler, H. and Alavi, B. (1998).
\newblock Development of improved design procedures for near-fault ground
  motions.
\newblock In {\em SMIP98 Seminar on Utilization of Strong-Motion Data}, pages
  21--41.

\bibitem[Langhaar, 1951]{Langhaar1951}
Langhaar, H.~L. (1951).
\newblock {\em Dimensional analysis and theory of models}.
\newblock Wiley.

\bibitem[Loh \textit{et~al.}, 2000]{LohLeeWuPeng2000}
Loh, C.-H., Lee, Z.-K., Wu, T.-C., and Peng, S.-Y. (2000).
\newblock Ground motion characteristics of the {C}hi-{C}hi earthquake of 21
  {S}eptember 1999.
\newblock {\em Earthquake Engineering \& Structural Dynamics}, 29(6):867--897.

\bibitem[Ma \textit{et~al.}, 2001]{MaMoriLeeYu2001}
Ma, K.-F., Mori, J., Lee, S.-J., and Yu, S. (2001).
\newblock Spatial and temporal distribution of slip for the 1999 {C}hi-{C}hi,
  {T}aiwan, earthquake.
\newblock {\em Bulletin of the Seismological Society of America},
  91(5):1069--1087.

\bibitem[Makris, 1997]{Makris1997}
Makris, N. (1997).
\newblock Rigidity--plasticity--viscosity: {C}an electrorheological dampers
  protect base-isolated structures from near-source ground motions?
\newblock {\em Earthquake Engineering \& Structural Dynamics}, 26(5):571--591.

\bibitem[Makris and Black, 2004a]{MakrisBlack2004Dimensional_a}
Makris, N. and Black, C.~J. (2004a).
\newblock {\noopsort{A}{Dimensional}} analysis of rigid-plastic and
  elastoplastic structures under pulse-type excitations.
\newblock {\em Journal of Engineering Mechanics}, 130(9):1006--1018.

\bibitem[Makris and Black, 2004b]{MakrisBlack2004Dimensional_b}
Makris, N. and Black, C.~J. (2004b).
\newblock {\noopsort{B}{Dimensional}} analysis of bilinear oscillators under
  pulse-type excitations.
\newblock {\em Journal of Engineering Mechanics}, 130(9):1019--1031.

\bibitem[Makris and Chang, 2000]{MakrisChang2000}
Makris, N. and Chang, S.-P. (2000).
\newblock Response of damped oscillators to cycloidal pulses.
\newblock {\em Journal of Engineering Mechanics}, 126(2):123--131.

\bibitem[Makris and Psychogios, 2006]{MakrisPsychogios2006}
Makris, N. and Psychogios, T. (2006).
\newblock Dimensional response analysis of yielding structures with first-mode
  dominated response.
\newblock {\em Earthquake Engineering \& Structural Dynamics},
  35(10):1203--1224.

\bibitem[MATLAB, 2017]{MATLAB2017}
MATLAB (2017).
\newblock {\em Version 9.2 (R2017a)}.
\newblock The MathWorks Inc., Natick, Massachusetts.

\bibitem[Mavroeidis and Papageorgiou, 2003]{MavroeidisPapageorgiou2003}
Mavroeidis, G.~P. and Papageorgiou, A.~S. (2003).
\newblock A mathematical representation of near-fault ground motions.
\newblock {\em Bulletin of the Seismological Society of America},
  93(3):1099--1131.

\bibitem[Newmark, 1965]{Newmark1965}
Newmark, N.~M. (1965).
\newblock Effects of earthquakes on dams and embankments.
\newblock {\em Geotechnique}, 15(2):139--160.

\bibitem[Rathje \textit{et~al.}, 1998]{RathjeAbrahamsonBray1998}
Rathje, E.~M., Abrahamson, N.~A., and Bray, J.~D. (1998).
\newblock Simplified frequency content estimates of earthquake ground motions.
\newblock {\em Journal of Geotechnical and Goenvironmental Engineering},
  124(2):150--159.

\bibitem[Ricker, 1943]{Ricker1943}
Ricker, N. (1943).
\newblock Further developments in the wavelet theory of seismogram structure.
\newblock {\em Bulletin of the Seismological Society of America},
  33(3):197--228.

\bibitem[Ricker, 1944]{Ricker1944}
Ricker, N. (1944).
\newblock Wavelet functions and their polynomials.
\newblock {\em Geophysics}, 9(3):314--323.

\bibitem[Rodriguez-Marek, 2000]{RodriguezMarek2000}
Rodriguez-Marek, A. (2000).
\newblock {\em Near-fault seismic site response}.
\newblock PhD thesis, Univeristy of California, Berkeley.

\bibitem[Sekiguchi and Iwata, 2002]{SekiguchiIwata2002}
Sekiguchi, H. and Iwata, T. (2002).
\newblock Rupture process of the 1999 {K}ocaeli, {T}urkey, earthquake estimated
  from strong-motion waveforms.
\newblock {\em Bulletin of the Seismological Society of America},
  92(1):300--311.

\bibitem[Somerville and Graves, 1993]{SomervilleGraves1993}
Somerville, P. and Graves, R. (1993).
\newblock Conditions that give rise to unusually large long period ground
  motions.
\newblock {\em The structural design of tall buildings}, 2(3):211--232.

\bibitem[Somerville, 1998]{Somerville1998}
Somerville, P.~G. (1998).
\newblock Development of an improved representation of near fault ground
  motions.
\newblock In {\em SMIP98 Seminar on Utilization of Strong-Motion Data},
  volume~15, page 1998.

\bibitem[Somerville, 2003]{Somerville2003}
Somerville, P.~G. (2003).
\newblock Magnitude scaling of the near fault rupture directivity pulse.
\newblock {\em Physics of the earth and planetary interiors},
  137(1-4):201--212.

\bibitem[UBC, 1997]{UBC1997}
UBC (1997).
\newblock {\em 1997 Uniform Building Code}.

\bibitem[Vassiliou and Makris, 2011]{VassiliouMakris2011}
Vassiliou, M.~F. and Makris, N. (2011).
\newblock Estimating time scales and length scales in pulselike earthquake
  acceleration records with wavelet analysis.
\newblock {\em Bulletin of the Seismological Society of America},
  101(2):596--618.

\bibitem[Veletsos and Newmark, 1960]{VeletsosNewmark1960}
Veletsos, A.~S. and Newmark, N.~M. (1960).
\newblock Effect of inelastic behavior on the response of simple systems to
  earthquake motions.
\newblock In {\em Proceedings of the 2$^{\text{nd}}$ World Conference on
  Earthquake Engineering}, pages 895--912.

\bibitem[Veletsos \textit{et~al.}, 1965]{VeletsosNewmarkChelapati1965}
Veletsos, A.~S., Newmark, N.~M., and Chelapati, C. (1965).
\newblock Deformation spectra for elastic and elastoplastic systems subjected
  to ground shock and earthquake motions.
\newblock In {\em Proceedings of the 3$^{\text{rd}}$ World Conference on
  Earthquake Engineering}, volume~2, pages 663--682.

\bibitem[Wang \textit{et~al.}, 2001]{WangChangChen2001}
Wang, W.-H., Chang, S.-H., and Chen, C.-H. (2001).
\newblock Fault slip inverted from surface displacements during the 1999
  {C}hi-{C}hi, {T}aiwan, earthquake.
\newblock {\em Bulletin of the Seismological Society of America},
  91(5):1167--1181.

\bibitem[Weisberg, 2005]{Weisberg2005}
Weisberg, S. (2005).
\newblock {\em Applied linear regression}, volume 528.
\newblock John Wiley \& Sons.

\end{thebibliography}

\clearpage

\begin{appendices}
\setcounter{table}{0}
\renewcommand{\thetable}{S\arabic{table}}
\section*{\Large{Supplement}}
\vspace{-0.5cm}

\subsection*{Tables of Strong Ground Motions}
\vspace{-0.5cm}

The list of pulse-like ground motions and their pulse characteristics are presented in Tables \ref{tab:TabS1}, \ref{tab:TabS2} and \ref{tab:TabS3}. For the database of pulses, records of strong ground motions from the Next Generation Attenuation (NGA) \citep{AnchetaDarraghStewartSeyhanSilvaChiouWooddellGravesKottkeBooreKishidaDonahue2013} database and the European Strong Motion Database \citep{AmbraseysSmitSigbjornssonSuhadolcMargaris2002} with the following characteristics were used:
\begin{itemize}
\item Moment magnitude, $M_\text{W} \geq$ 4.5, and
\vspace{-.2cm}
\item Epicentral distance, $D \leq$ 20 km
\end{itemize}
The records where categorized based on their fault mechanism, as provided by the descriptions of the NGA and the European Strong Motion databases.

\begin{table}[htbp]
\setlength{\tabcolsep}{1pt}
{\normalsize
{
\begin{tabular}{ll}
	$D$ [km] & : epicentral distance \\
	$M_\text{W}$ & : moment magnitude \\
	PI & : pulse indicator \\
    $e_a$ & : matching index for the extracted acceleration pulse \\
    $e_v$ & : matching index for the obtained (from the acceleration pulse) velocity pulse \\
	$a_p$ [$g$] & : amplitude of the acceleration pulse \\
	$v_p$ [m/s] & : corresponding velocity amplitude of the elementary wavelet \\
	$t_{\text{0}}$ [s] & : time instance when the pulse starts \\
	$T_p$ [s] & : period of the acceleration pulse \\
	$\phi$ [rad] &: phase of the elementary wavelet \\
	$\gamma$ &: parameter that controls the oscillatory character of the elementary wavelet \\
	$S_a$ [$g$] &: pseudoacceleration of an elastic SDoF oscillator \\
			 & The pulses that result in larger pseudoaccelerations of the two elastic SDoF oscillators  \\
			 & with $T_{\text{0}}=$ 0.5 and 2.5 s than those of recorded ground motions are shown in \textbf{bold} font

\end{tabular}}}
\label{tab:Legend}
\end{table}

\begin{landscape}

\begin{table}
\begin{center}
\caption{Near-source, pulse-like ground motions from \underline{\textbf{dip-slip normal faults}} and the parameters of the extracted pulses. }
\includegraphics[page=1, trim=0.7in 0.75in 0.7in 0.75in,clip, width=1.\linewidth]{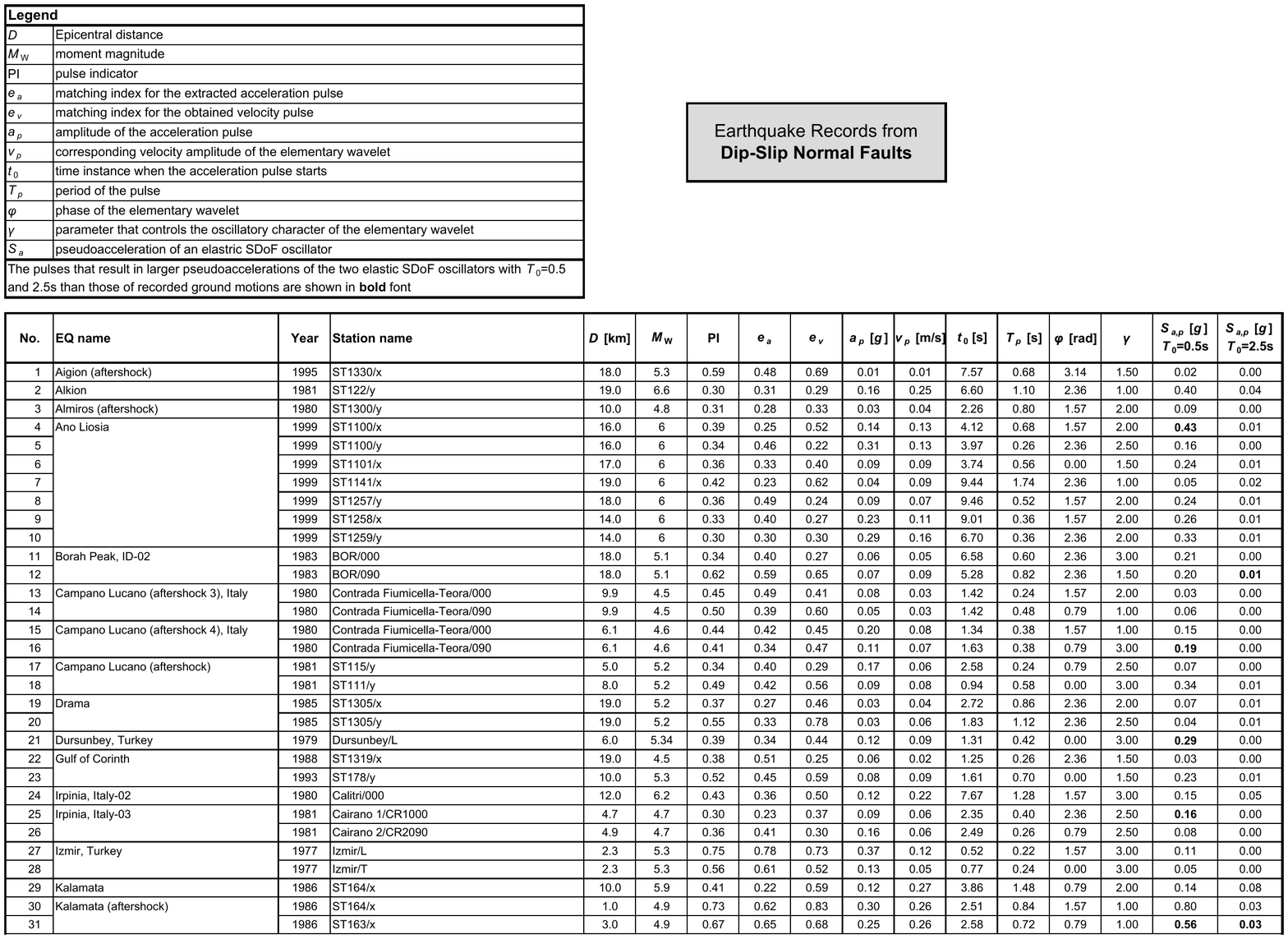}
\label{tab:TabS1}
\end{center}
\end{table}

\foreach \x in {2,...,3}
{%
\clearpage
\begin{table}
\begin{center}
\includegraphics[page=\x, trim=0.7in 0.75in 0.7in 0.75in,clip, width=1.\linewidth]{Normal_Records.pdf}
\end{center}
\end{table}
}

\begin{table}
\begin{center}
\caption{Near-source, pulse-like  ground motions from \underline{\textbf{dip-slip reverse faults}} and the parameters of the extracted pulses.}
\includegraphics[page=1, trim=0.7in 0.75in 0.7in 0.75in,clip, width=1.\linewidth]{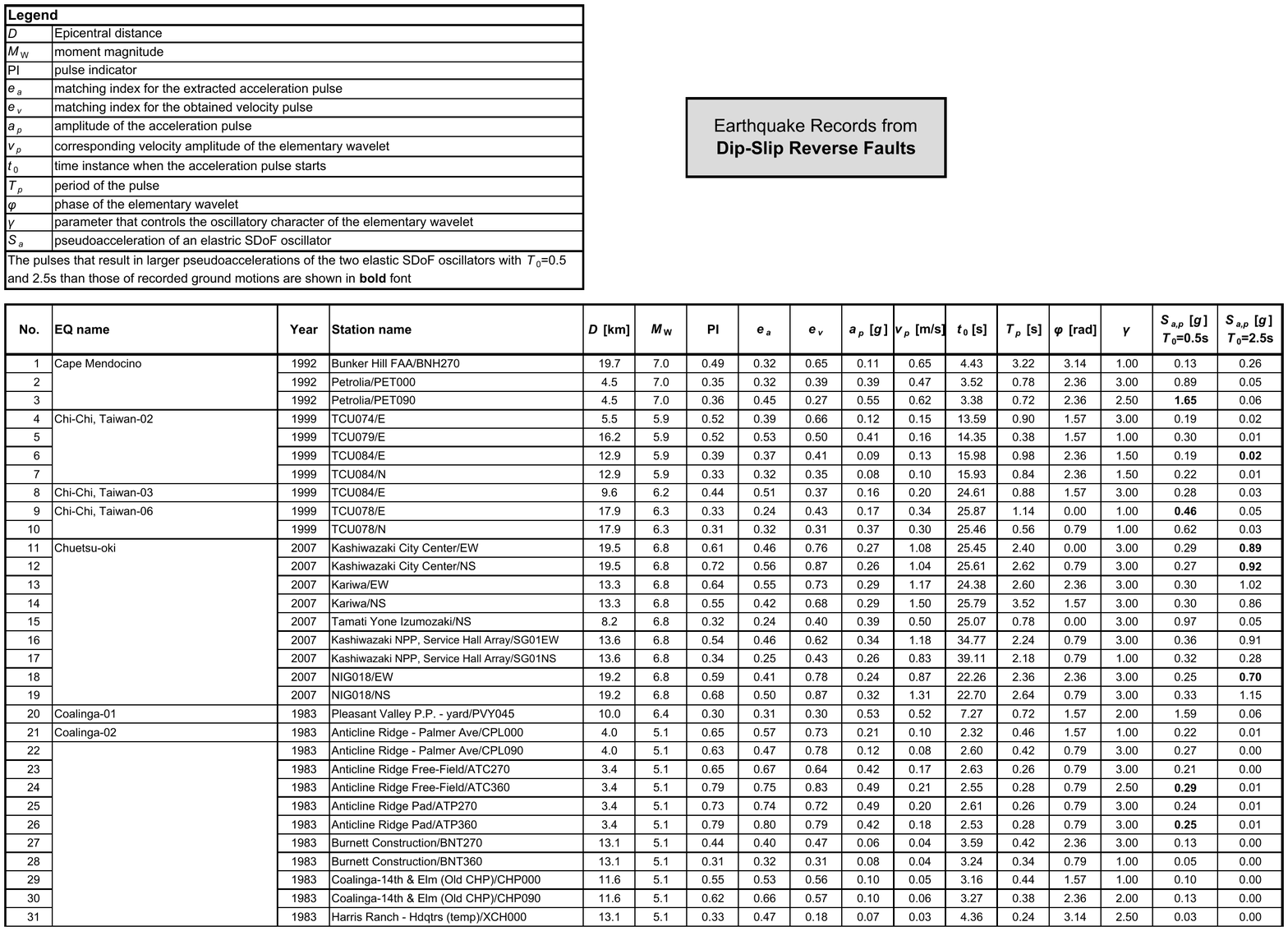}
\label{tab:TabS2}
\end{center}
\end{table}

\foreach \x in {2,...,5}
{%
\clearpage
\begin{table}
\begin{center}
\includegraphics[page=\x, trim=0.7in 0.75in 0.7in 0.75in,clip, width=1.\linewidth]{Reverse_Records.pdf}
\end{center}
\end{table}
}

\begin{table}
\begin{center}
\caption{Near-source, pulse-like  ground motions from \underline{\textbf{strike-slip faults}} and the parameters of the extracted pulses.}
\includegraphics[page=1, trim=0.7in 0.75in 0.7in 0.75in,clip, width=1.\linewidth]{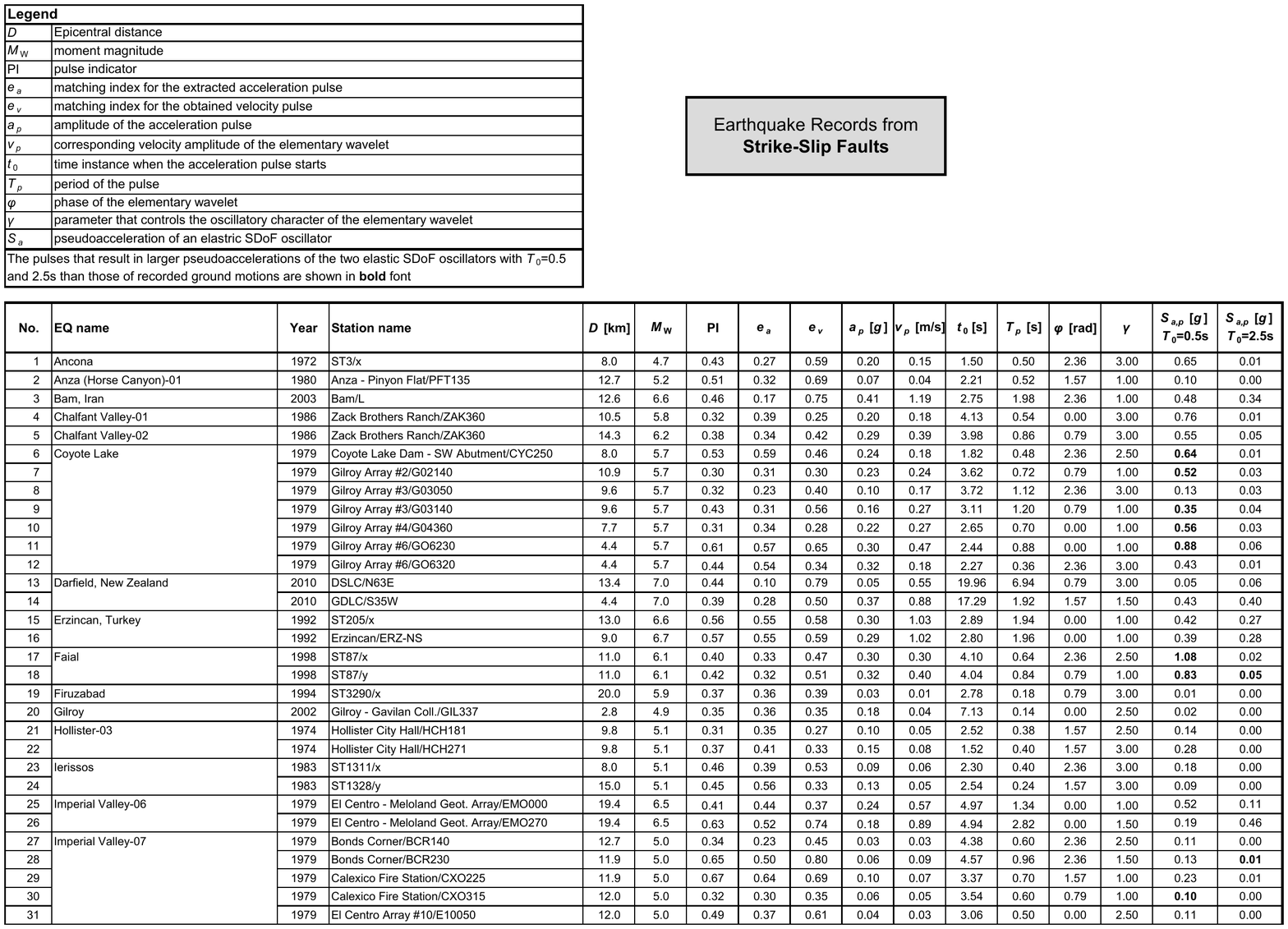}
\label{tab:TabS3}
\end{center}
\end{table}

\foreach \x in {2,...,3}
{%
\clearpage
\begin{table}
\begin{center}
\includegraphics[page=\x, trim=0.7in 0.75in 0.7in 0.75in,clip, width=1.\linewidth]{Strike-Slip_Records.pdf}
\end{center}
\end{table}
}
\end{landscape}
\end{appendices}

\end{document}